\documentclass[showpacs,aps,prd,reprint,superscriptaddress,nofootinbib,longbibliography,preprintnumbers]{revtex4-1}
\usepackage[colorlinks=true, pdfstartview=FitV, linkcolor=magenta,citecolor=blue, urlcolor=magenta,
bookmarks=true, bookmarksnumbered=true, breaklinks]{hyperref}
\usepackage[dvipdfmx]{graphicx}

\usepackage{here}
\usepackage{url}
\usepackage{amsmath,amssymb,bm,color,longtable,mathrsfs,amsfonts,slashed}

\begin{document}
\begin{flushright}
\end{flushright}

\title{FRG analysis of dense two-color QCD within the linear sigma model}

\author{Gergely Fej\H{o}s}
\affiliation{Institute of Physics and Astronomy, E\"otv\"os University, 1117 Budapest, Hungary,}
\affiliation{RIKEN Center for Interdisciplinary Theoretical and Mathematical Sciences (iTHEMS), Wako, Saitama 351-0198, Japan}
\author{Daiki Suenaga}
\affiliation{Kobayashi-Maskawa Institute for the Origin of Particles and the Universe, Nagoya University, Nagoya 464-8602, Japan}
\affiliation{Research Center for Nuclear Physics, Osaka University, Ibaraki 567-0048, Japan}

\date{\today}

\begin{abstract}
We investigate the phase structure, hadron masses, and topological susceptibility in the two-flavor and two-color QCD (QC$_2$D) medium, particularly focusing on the $U(1)_A$ axial anomaly effects. To this end, we employ the linear sigma model, and hadron fluctuations are incorporated through the functional renormalization group method. We establish in detail an effective potential that respects symmetries of QC$_2$D at finite quark chemical potential, $\mu_q$: $SU(2)_L\times SU(2)_R$ chiral, $U(1)$ baryon-number, parity and time-reversal symmetries. We find that the $U(1)_A$ anomaly couplings for mesons at finite temperature are enhanced with increasing $\mu_q$, while that of the baryons are not too sensitive to $\mu_q$. Despite the anomaly enhancement, we find that the topological susceptibility at larger $\mu_q$ is always suppressed regardless of the temperature, following chiral restoration. We also find that mass degeneracies of the chiral partners are well realized at higher temperatures and densities by the chiral restoration. Our findings are expected to provide useful information on properties of the $U(1)_A$ anomaly in medium for sign-problem-free lattice simulations of QC$_2$D.

\end{abstract}


\maketitle

\section{Introduction}
\label{sec:Introduction}

Quantum chromodynamics (QCD) at finite temperature and density is expected to exhibit a rich phase structure, including, e.g., chiral-symmetry restoration and color deconfinement. Hot QCD medium has been explored experimentally by means of heavy-ion collisions at, e.g., CERN and RHIC, where important properties of the quark-gluon plasma have been unveiled~\cite{Busza:2018rrf}. Meanwhile, the possibility of examining cold and dense medium has recently been opened up due to neutron star observations, including detection of gravitational waves~\cite{Baym:2017whm}. These experiments and observations are capable of providing clues, which may shed light on QCD related questions that cannot be achieved in vacuum.

One of the most powerful tools to explore QCD properties is lattice simulations, which are first-principle Monte-Carlo simulations done numerically on a discretized spacetime lattice. These {\it numerical experiments} work straightforwardly in the vacuum and at finite temperature, but not when a quark chemical potential $\mu_q$ is included. At finite $\mu_q$, the notorious {\it sign problem} prevents the use of Monte Carlo simulations, consequently, lattice QCD essentially breaks down at finite density.~\cite{Aarts:2015tyj,Nagata:2021ugx}. 

The sign problem appears since the fermion determinant becomes imaginary at finite density. This imaginary part disappears for two-color QCD (QC$_2$D), because of the pseudoreality of the $SU(2)$ group~\cite{Muroya:2003qs}. Moreover, in QC$_2$D it turns out that the fermion determinant with an even number of flavors is positive definite, which enables the evaluation of ensembles. Hence, QC$_2$D with $N_f=2$ offers the simplest and one of the most useful testing grounds to perform lattice simulations at finite $\mu_q$. So far, many QC$_2$D lattice simulations have been implemented~\cite{Boz:2019enj,Buividovich:2020dks,Astrakhantsev:2020tdl,Iida:2024irv,Braguta:2023yhd}, which motivated theoretical analyses to provide these {\it numerical experiments} with qualitative and predictive understandings~\cite{Kogut:2000ek,Lenaghan:2001sd,Ratti:2004ra,Sun:2007fc,Brauner:2009gu,Harada:2010vy,Strodthoff:2011tz,Imai:2012hr,Duarte:2015ppa,Contant:2019lwf,Suenaga:2019jjv,Khunjua:2020xws,Suenaga:2022uqn,Acharyya:2024pqj}. We note that an important feature of QC$_2$D is that the pseudoreality of $SU(2)$ leads to the extension of $SU(2)_L\times SU(2)_R$ chiral symmetry to the so-called Pauli-G\"{u}rsey $SU(4)$ symmetry~\cite{Pauli:1957voo,Gursey:1958fzy}.

In QC$_2$D, diquarks made of two quarks are counted as color-singlet hadrons. Since they are bosons, at low temperature $0^+$ diquarks can form a Bose-Einstein condensate (BEC)~\cite{Suenaga:2025sln}. This BEC phase is referred to as the {\it baryon superfluid phase}, reflecting its color-singlet nature. Meanwhile, the chiral broken phase without any diquark condensate is called the {\it hadronic phase}.

Recent QC$_2$D lattice simulations with a rather heavy pion mass ($m_\pi\sim 738$ MeV), at sufficiently low temperature ($T\sim40$ MeV) implied that the lowest excitation mode carrying negative parity is the iso-singlet ($I=0$) hadron in the superfluid phase, while the pion mass increases monotonically in this phase~\cite{Iida:2026dmw}. This peculiar mass ordering implies the importance of parity partners in the low-energy regime of dense QC$_2$D. Motivated by this fact, investigations of QC$_2$D properties based on the linear sigma model (LSM), with linear representation of the dynamical fields, have been under extensive study~\cite{Suenaga:2019jjv,Kawaguchi:2023olk,Suenaga:2023xwa,Kawaguchi:2024iaw,Fejos:2025nvd,Fejos:2025oxi,Suenaga:2025sln} (for a review see Ref.~\cite{Suenaga:2025sln}).

As long as we stick to zero temperature, a mean-field analysis is reasonably adopted, while exploration in a finite-$T$ system requires the inclusion of thermal fluctuations. In Refs.~\cite{Fejos:2025nvd,Fejos:2025oxi}, we adopted the functional renormalization group (FRG) method to take into account hadronic fluctuations within the LSM to access the hot medium. In these studies, possible chiral restoration patterns with Pauli-G\"{u}rsey $SU(4)$ symmetry and an enhancement of the $U(1)_A$ axial anomaly effects for the hadrons in hot medium were found. We note that the anomaly enhancement at finite temperature and density based on the FRG method was also predicted in three-color QCD~\cite{Fejos:2016hbp,Fejos:2017kpq,Fejos:2021yod}.

In the present work, we apply the FRG method to the finite $\mu_q$ system at various temperatures, to explore a broader range of QC$_2$D medium. We study the phase structure, hadron mass spectrum and topological susceptibility in the medium, particularly focusing on the $U(1)_A$ anomaly effects and the chiral restoration. We also discuss the possibility of anomaly enhancement in medium. 

The paper is organized as follows. In Sec.~\ref{sec:LSM}, a variant of the LSM that respects all the symmetries of QC$_2$D at finite $\mu_q$ is introduced, while in Sec.~\ref{sec:FlowEQ} properties of the FRG flow equations are briefly explained. In Sec.~\ref{sec:Numerical} we present our main results on the phase structure, hadron mass spectrum and the topological susceptibility. The enhancement of the $U(1)_A$ anomaly effects for the mesons is also shown in this section. Finally, Sec.~\ref{sec:Conclusions} is devoted to conclusions.

\section{The LSM at finite $\mu_q$}
\label{sec:LSM}

First, we introduce the LSM that respects all the symmetries of QC$_2$D with finite $\mu_q$, which will play a central role in the upcoming FRG analysis.

The pseudoreality of $SU(2)$ color theory, ${\bm 2} \cong \bar{\bm 2}$, allows us to treat mesons and diquark baryons on an equal footing, i.e., they belong to the same multiplet. This leads to an extension of ordinary $SU(2)_L \times SU(2)_R$ chiral symmetry to the Pauli-G\"{u}rsey $SU(4)$ symmetry~\cite{Kogut:1999iv,Kogut:2000ek}. As long as the quark mass, $m_q$, and the chemical potential, $\mu_q$, are treated perturbatively, the LSM is constructed based on $SU(4)$, with a linear representation of the quark operators. 
Within this treatment, a fundamental building block for constructing the LSM Lagrangian is the following $4\times4$ matrix~\cite{Suenaga:2025sln}:
\begin{eqnarray}
\Sigma &=& \sum_{a=0}^5({\cal S}^a-i{\cal P}^a)X^a E\ , \label{SigmaDef}
\end{eqnarray}
where ${\cal S}^{a}$ and ${\cal P}^a$ are related to the particle basis as
\begin{eqnarray}
\sigma = {\cal S}^{0} \, , \  a_0^{\pm} = \frac{{\cal S}^{1}\mp i{\cal S}^2}{\sqrt{2}} \, , \ \eta = {\cal P}^{0}  \,  ,\  \pi^{\pm} = \frac{{\cal P}^{1}\mp i{\cal P}^2}{\sqrt{2}}\, ,
\end{eqnarray}
and
\begin{eqnarray}
B (\bar{B})= \frac{{\cal P}^5 \mp i{\cal P}^4}{\sqrt{2}}\ , \ \ B' (\bar{B}')= \frac{{\cal S}^5 \mp i{\cal S}^4}{\sqrt{2}}\ .
\end{eqnarray}
The $4\times4 $ matrices $X^a$ ($a=1$ - $5$) are generators of $SU(4)/Sp(4)$:
\begin{eqnarray}
&&X^{a=1-3} = \frac{1}{2\sqrt{2}}\left(
\begin{array}{cc}
\tau_f^a & 0 \\
0 &(\tau_f^a)^T \\
\end{array}
\right)\  , \nonumber\\
&& X^{a=4,5} = \frac{1}{2\sqrt{2}}\left(
\begin{array}{cc}
0& D_f^a \\
(D_f^a)^\dagger & 0 \\
\end{array}
\right) \ , \label{GeneratorX}
\end{eqnarray}
with $D_f^4=\tau_f^2$ and $D_f^5=i\tau_f^2$ ($\tau_f^{a}$ are the Pauli matrices acting in flavor space), and $X^{a=0} = {\bm 1}/(2\sqrt{2})$. In Eq.~(\ref{SigmaDef}), $E$ is the $4\times4$ symplectic matrix,
\begin{eqnarray}
E = \left(
\begin{array}{cc}
 {\bm 0} & {\bm 1}\\
 -{\bm 1} & {\bm 0} \\
 \end{array}
 \right)\ .
\end{eqnarray}
With Eq.~(\ref{SigmaDef}), the kinetic term takes the simple form of
\begin{eqnarray}
{\cal L}_{\rm kin} = {\rm Tr}[D_\mu\Sigma^\dagger D^\mu\Sigma]\ , \label{LKin}
\end{eqnarray}
where $D_\mu\Sigma = \partial_\mu\Sigma-i{\cal V}_\mu\Sigma-i\Sigma{\cal V}^T_\mu$ with a vector external field ${\cal V}_\mu = \mu_q\delta_{\mu0}J$ and $J={\rm diag}(1,1,-1,-1)$.

\begin{table}[t]
\begin{center}
  \begin{tabular}{c|ccc}  \hline\hline
  & $\Sigma_M$ & $\Sigma_{B_R}$ & $\Sigma_{B_L}$ \\ \hline
Chiral & $\Sigma_M \to g_L\Sigma_M g_R^\dagger $& $\Sigma_{B_R} \to \Sigma_{B_R}$ & $\Sigma_{B_L} \to \Sigma_{B_L}$ \\
$U(1)_B$ & $\Sigma_M \to \Sigma_M $& $\Sigma_{B_R} \to g_V\Sigma_{B_R}$ & $\Sigma_{B_L} \to g_V\Sigma_{B_L}$ \\
$U(1)_A$ & $\Sigma_M \to (g_A^\dagger)^2 \Sigma_M $& $\Sigma_{B_R} \to g^2_A\Sigma_{B_R}$ & $\Sigma_{B_L} \to (g_A^\dagger)^2\Sigma_{B_L}$ \\
\hline \hline
 \end{tabular}
\caption{ $SU(2)_L\times SU(2)_R$ chiral, $U(1)$ baryon and $U(1)$ axial transformation laws of $\Sigma_M$, $\Sigma_{B_R}$ and $\Sigma_{B_L}$.}
\label{tab:Transform}
\end{center}
\end{table}

As long as we stick to perturbative analyses, even at finite $\mu_q$, it is meaningful to construct the interaction Lagrangian in terms of the field $\Sigma$, based on the Pauli-G\"{u}rsey $SU(4)$ symmetry. However, when we employ non-perturbative methods, such as the FRG at finite density, it is necessary to take into account the fact that in the presence of $\mu_q$, which obviously separates the meson and baryon sectors, the symmetry of the system gets modified. That is, we must construct the interacting Lagrangian not on the basis of the full $SU(4)$ symmetry, but rather on that of its subgroup, $SU(2)_L \times SU(2)_R \times U(1)_B$. If we used the original, $SU(4)$ symmetric Lagrangian, we would immediately find that the renormalization group flow equations do not close under such interactions at finite chemical potential. For this reason, we explicitly decompose $\Sigma$ into the meson and baryon parts:
\begin{eqnarray}
\Sigma = \frac{1}{2\sqrt{2}}\left(
\begin{array}{cc}
i\Sigma_{B_R}& \Sigma_M^\dagger \\
-\Sigma_M^* & -i\Sigma_{B_L}^\dagger \\
\end{array}
\right)\ ,
\label{SigmaEx}
\end{eqnarray}
with
\begin{eqnarray}
\Sigma_{B_R} &\equiv& \sum_{a=4}^5\big({\cal P}^a + i{\cal S}^a\big)D_f^a \ , \nonumber\\
\Sigma_{B_L} &\equiv& \sum_{a=4}^5\big({\cal P}^a - i{\cal S}^a\big)D_f^a \ , \nonumber\\
\Sigma_M &\equiv& \sum_{a=0}^3\big({\cal S}^a + i{\cal P}^a\big)\tau_f^a\ .
\end{eqnarray}
The $SU(2)_L\times SU(2)_R$ chiral, $U(1)$ baryon and $U(1)$ axial transformation laws of those $2\times2$ matrices are summarized in Table~\ref{tab:Transform}.

In terms of $\Sigma_M$, $\Sigma_{B_R}$ and $\Sigma_{B_L}$, the kinetic term~(\ref{LKin}) takes the form of
\begin{eqnarray}
{\cal L}_{\rm kin}  &=&\frac{1}{4}{\rm tr}[\partial_\mu\Sigma_M^\dagger \partial^\mu\Sigma_M] \nonumber\\
&& +\frac{1}{8}{\rm tr}\left[D_\mu\Sigma_{B_R}^\dagger D^\mu\Sigma_{B_R}+D_\mu\Sigma_{B_L}^\dagger D^\mu\Sigma_{B_L}\right] \,  ,  \label{LKin2}
\end{eqnarray}
where $D_\mu \Sigma_{B_{R/L}} =( \partial_\mu -2i\mu_q\delta_{\mu0})\Sigma_{B_{R/L}}$. At finite $\mu_q$, the remaining symmetry is $SU(2)_L\times SU(2)_R$ chiral, $U(1)$ baryon-number, parity and time-reversal symmetries. Hence, when taking into account interactions up to ${\cal O}(\Sigma^4)$, from Table~\ref{tab:Transform} one obtains the following potential:
\begin{eqnarray}
V = V_{\rm AF} + V_{\rm F}\ , \label{VSum}
\end{eqnarray}
with
\begin{widetext}
\begin{eqnarray}
V_{\rm AF} &=& m_M^2{\rm tr}\left[\Sigma_M^\dagger\Sigma_M\right] + m^2_B{\rm tr}\left[\Sigma_{B_R}^\dagger\Sigma_{B_R} + \Sigma_{B_L}^\dagger\Sigma_{B_L}\right] +\lambda_{M1} \left({\rm tr}\left[\Sigma_M^\dagger\Sigma_M\right]\right)^2 \nonumber\\
&+& \lambda_{M2} {\rm tr}\left[\left(\Sigma_M^\dagger\Sigma_M - \frac{1}{2}{\rm tr}[\Sigma_M^\dagger\Sigma_M]\right)^2\right] +\lambda_{B1} \left( {\rm tr}\left[\Sigma_{B_R}^\dagger\Sigma_{B_R} + \Sigma_{B_L}^\dagger\Sigma_{B_L}\right] \right)^2  \nonumber\\
&+& \lambda_{B2} \left( {\rm tr}\left[\Sigma_{B_R}^\dagger\Sigma_{B_R} - \Sigma_{B_L}^\dagger\Sigma_{B_L}\right] \right)^2+ \gamma_1{\rm tr}\left[\Sigma_M^\dagger\Sigma_M\right]{\rm tr}\left[\Sigma_{B_R}^\dagger\Sigma_{B_R} + \Sigma_{B_L}^\dagger\Sigma_{B_L}\right] \nonumber\\
&+& \gamma_2\left({\rm tr}\left[\Sigma_{B_R}^\dagger\Sigma_{B_L} \right]{\rm det}\Sigma^\dagger_M + {\rm tr}\left[ \Sigma_{B_L}^\dagger\Sigma_{B_R}\right]{\rm det}\Sigma_M \right) \ , \label{VAFGeneral1}
\end{eqnarray}
and
\begin{eqnarray}
V_{\rm A}&=& a_{M}\left({\rm det}\Sigma_M+{\rm det}\Sigma_M^\dagger\right) + a_{B}{\rm tr}\left[\Sigma_{B_R}^\dagger\Sigma_{B_L} + \Sigma_{B_L}^\dagger\Sigma_{B_R}\right] + c_{M1}\left(({\rm det}\Sigma_M)^2+({\rm det}\Sigma_M^\dagger)^2\right) \nonumber\\
&+& c_{M2}{\rm tr}\left[\Sigma_M^\dagger\Sigma_M\right]\left({\rm det}\Sigma_M+{\rm det}\Sigma_M^\dagger\right) + c_{B1} \left(({\rm tr}[\Sigma_{B_R}^\dagger\Sigma_{B_L}])^2 + ( {\rm tr}[\Sigma_{B_L}^\dagger\Sigma_{B_R}])^2\right) \ , \nonumber\\
&+&  c_{B2} {\rm tr}\left[\Sigma_{B_R}^\dagger\Sigma_{B_R} + \Sigma_{B_L}^\dagger\Sigma_{B_L}\right]  {\rm tr}\left[\Sigma_{B_R}^\dagger\Sigma_{B_L} + \Sigma_{B_L}^\dagger\Sigma_{B_R}\right]  +  d_1{\rm tr}\left[\Sigma_M^\dagger\Sigma_M\right]{\rm tr}\left[\Sigma_{B_R}^\dagger\Sigma_{B_L} + \Sigma_{B_L}^\dagger\Sigma_{B_R}\right] \nonumber\\
 &+& d_2 {\rm tr}\left[\Sigma_{B_R}^\dagger\Sigma_{B_R} + \Sigma_{B_L}^\dagger\Sigma_{B_L}\right]  \left({\rm det}\Sigma_M+{\rm det}\Sigma_M^\dagger\right) + d_3 \left({\rm tr}\left[\Sigma_{B_R}^\dagger\Sigma_{B_L} \right]{\rm det}\Sigma_M + {\rm tr}\left[ \Sigma_{B_L}^\dagger\Sigma_{B_R}\right]{\rm det}\Sigma^\dagger_M \right) \ .
 \label{VAGeneral1} 
\end{eqnarray}
\end{widetext}
Here, $V_{\rm AF}$ and $V_{\rm A}$ stand for non-anomalous and anomalous contributions in terms of $U(1)_A$, respectively. In the present work, explicit chiral-symmetry breaking is introduced via the following source term:
\begin{eqnarray}
{\cal L}_{\rm ex} = \bar{c}m_q{\rm tr}[E^T\Sigma + \Sigma^\dagger E]\ . \label{Vex}
\end{eqnarray}
Using Eqs.~(\ref{LKin2}),~(\ref{VSum}) and~(\ref{Vex}), the general Lagrangian at the classical level at finite $\mu_q$ is derived as
\begin{eqnarray}
{\cal L}_{\rm cl} \equiv {\cal L}_{\rm kin} + {\cal L}_{\rm ex}  -V  \ . \label{LClassical}
\end{eqnarray}

We note that FRG analyses of QC$_2$D medium were done in Refs.~\cite{Strodthoff:2011tz,Strodthoff:2013cua}. In these references, however, only $\sigma$, $\pi$ and $B(\bar{B})$ modes are taken into account, which indeed form a closed set, unless the $U(1)_A$ property is concerned. We wish to highlight that in our present study, motivated by the importance of $\eta$ and $B'(\bar{B}')$ modes, as the lattice data suggests, we have also incorporated all the parity partners. This extension allows us to examine the anomalous breaking of $U(1)_A$ in the medium consistently.

\section{Flow equations}
\label{sec:FlowEQ}

\subsection{General properties}
\label{sec:FlowGeneral}

Although the Lagrangian~(\ref{LClassical}) captures general symmetry properties of dense QC$_2$D within the linear representation, it describes a classical theory and thus quantum fluctuations are not taken into account at this level. One promising way to include the latter effects non-perturbatively is to solve renormalization group flow equations, e.g., via the FRG method~\cite{Berges:2000ew,Dupuis:2020fhh}. 

When employing the FRG framework, we focus on a scale-dependent effective action $\Gamma_k$ that incorporates fluctuations with momenta $p\gtrsim k$, where $k$ plays the role of a scale separation variable. At large $k$, practically at some ultraviolet (UV) cutoff, $k\to \Lambda$, none of the fluctuations are switched on, therefore, the effective action is reduced to a classical one. In our present case, this classical theory corresponds to the action based on Eq.~(\ref{LClassical}). The full quantum action, with all the fluctuations included, is evaluated at the infrared (IR) scale, $k\to 0$. 

The renormalization group flow of the $k$-dependent effective action $\Gamma_{k}$ is determined by solving the Wetterich equation~\cite{Wetterich:1992yh}
\begin{eqnarray}
\partial_k \Gamma_k = \frac12 {\rm Tr} \Big\{ \partial_k R_k [\Gamma_k''+R_k]^{-1}\Big\} \ ,\label{Wet1}
\end{eqnarray}
where $R_k$ is a regulator satisfying $R_{k\to\infty}=\infty$ and $R_{k\to0}=0$. For $R_k$, we adopt the 3D Litim regulator~\cite{Litim:2001up}, which in Fourier space reads
\begin{eqnarray}
 R_k = (k^2-{\bm p}^2)\Theta(k^2-{\bm p}^2), \label{LitimR}
 \end{eqnarray}
since we aim to delineate properties of QC$_2$D in medium that breaks Lorentz symmetry. In Eq.~(\ref{Wet1}), ``Tr'' stands for a trace operator acting on spacetime and hadron fields. The double dashes in $\Gamma_k''$ represent the second derivatives of $\Gamma_k$ with respect to the $12$ hadron fields: ${\cal S}^a$ and ${\cal P}^a$ ($a=0$ - $5$).

In principle, solving the flow equation~(\ref{Wet1}) yields a complicated non-local effective action with arbitrary interactions respecting symmetries of QC$_2$D medium. In the present exploratory study, we adopt the local potential approximation (LPA), and assume that even at the quantum level the effective action takes the same form as the classical one:~(\ref{LClassical}): $\Gamma_k = \int d^4x\,  {\cal L}_k$ with
\begin{eqnarray}
{\cal L}_k = {\cal L}_{\rm kin} + {\cal L}_{\rm ex}  -V_k\ , \label{LatK}
\end{eqnarray}
where all the coefficients in $V_k$ are now $k$-dependent. We note that $V_{\rm ex}$ does not flow as this term is always regarded as an external source, i.e., it is linear in the dynamical field, therefore cannot be modified throughout the RG flow.

\subsection{Right-hand sides of flow equations}
\label{sec:EvaluateFlow}
 
In this subsection, we briefly explain how the flow equations for individual couplings are derived. From here onwards, we adopt the imaginary-time formalism~\cite{Kapusta:2006pm} and use the shorthand notation of $\int_{\bm p} = \int d^3p/(2\pi)^3$. 
When evaluating the right-hand side (r.h.s.) of the flow equation, note that, the $\tilde{\partial}_k$ derivative by definition only acts on the regulator $R_k({\bm p})$, so in what follows we will omit the subscript $k$ for the potential and couplings as a shorthand notation.

From our assumption of Eq.~(\ref{LatK}), the r.h.s. of Eq.~(\ref{Wet1}), after dividing by the volume read
\begin{eqnarray}
({\rm r.h.s.}) =  -\frac{\tilde{\partial}_k}{2}T\sum_n\int_{\bm p}{\rm tr} \, {\rm ln}\Big[\Omega_n^2+{\bm p}^2 +R_k({\bm p}) + V''\Big]\, , \nonumber\\
\label{RHSFlow}
\end{eqnarray}
with
\begin{eqnarray}
\tilde{\partial}_k \equiv \partial_k R_k \frac{\partial_k}{\partial R_k}\ .
\end{eqnarray}
The trace operator ``tr'' in Eq.~(\ref{RHSFlow}) acts on the $12\times12$ hadron-field matrix. As for the Matsubara-frequency matrix $\Omega_n^2$, one can easily see that
\begin{eqnarray}
&&(\Omega_n^2)_{{\cal S}^0{\cal S}^0} = (\Omega_n^2)_{{\cal S}^1{\cal S}^1} =(\Omega_n^2)_{{\cal S}^2{\cal S}^2} = (\Omega_n^2)_{{\cal S}^3{\cal S}^3}= \omega_n^2\ , \nonumber\\
&& (\Omega_n^2)_{{\cal P}^0{\cal P}^0} = (\Omega_n^2)_{{\cal P}^1{\cal P}^1} =(\Omega_n^2)_{{\cal P}^2{\cal P}^2} = (\Omega_n^2)_{{\cal P}^3{\cal P}^3}= \omega_n^2\ , \nonumber\\
\end{eqnarray}
while
\begin{eqnarray}
&& (\Omega_n^2)_{{\cal S}^4{\cal S}^4} = (\Omega_n^2)_{{\cal S}^5{\cal S}^5} = \omega_n^2-4\mu_q^2\ , \nonumber\\
&& (\Omega_n^2)_{{\cal S}^4{\cal S}^5} =- (\Omega_n^2)_{{\cal S}^5{\cal S}^4} = 4\mu_q\omega_n\ , \nonumber\\
&& (\Omega_n^2)_{{\cal P}^4{\cal P}^4} = (\Omega_n^2)_{{\cal P}^5{\cal P}^5} = \omega_n^2-4\mu_q^2\ , \nonumber\\
&& (\Omega_n^2)_{{\cal P}^4{\cal P}^5} =- (\Omega_n^2)_{{\cal P}^5{\cal P}^4} = 4\mu_q\omega_n\ ,
\end{eqnarray}
and the remaining elements are vanishing, with $\omega_n=2n\pi T$ ($n\in \mathbb{Z}$).

In order to derive the flow equations for the $17$ individual couplings (i.e., $m_{M}^2$, $m_{B}^2$, $\cdots$), one expands Eq.~(\ref{RHSFlow}) up to ${\cal O}(\phi^4)$ (here $\phi$ stands for $12$ hadron fields collectively: $\phi\ni \{{\cal S}^a, {\cal P}^a\}$). This expansion reads
\begin{eqnarray}
({\rm r.h.s.})&\approx& -\frac{\tilde{\partial}_k}{2}T\sum_n\int_{\bm p}{\rm tr}\Big[D_{0,k}^{-1}\Big] \nonumber\\
&& -  \frac{\tilde{\partial}_k}{2}T\sum_n\int_{\bm p}{\rm tr}\Big[D_{0,k}V_{(2)}''\Big] \nonumber\\
&& +\frac{\tilde{\partial}_k}{4}T\sum_n\int_{\bm p}{\rm tr}\Big[D_{0,k}V_{(2)}''D_{0,k}V_{(2)}''\Big]\ . \label{RHSExpand}
\end{eqnarray}
Since the potential $V$ contains two-point and four-point interactions, taking $V''$ leaves ${\cal O}(\phi^0)$ and ${\cal O}(\phi^2)$ contributions. Hence, we have separated those two as $V'' = V''_{(0)} + V''_{(2)}$ in Eq.~(\ref{RHSExpand}). With this notation, the leading-order inverse propagator $D^{-1}_{0,k}$ is defined to be
\begin{eqnarray}
D_{0,k}^{-1} = \Omega_n^2+{\bm p}^2+R_k({\bm p})+V_{(0)}''\ ,
\end{eqnarray}
which is diagonal in the ${\cal S}^{a=0-3}$ and ${\cal P}^{a=0-3}$ sectors. Therefore,
\begin{eqnarray}
&&D_{0,k}^{{\cal S}^0{\cal S}^0} = D_{0,k}^{{\cal P}^1{\cal P}^1} = D_{0,k}^{{\cal P}^2{\cal P}^2} = D_{0,k}^{{\cal P}^3{\cal P}^3} \nonumber\\
  &=& -\frac{1}{2E_{k,M+}}\left(\frac{1}{i\omega_n-E_{k,M+}} - \frac{1}{i\omega_n+E_{k,M+}} \right)\ , \nonumber\\ \label{DMeson1}
\end{eqnarray}
\begin{eqnarray}
&&D_{0,k}^{{\cal P}^0{\cal P}^0} = D_{0,k}^{{\cal S}^1{\cal S}^1} = D_{0,k}^{{\cal S}^2{\cal S}^2} = D_{0,k}^{{\cal S}^3{\cal S}^3} \nonumber\\
  &=& -\frac{1}{2E_{k,M-}}\left(\frac{1}{i\omega_n-E_{k,M-}} - \frac{1}{i\omega_n+E_{k,M-}} \right)\ , \nonumber\\ \label{DMeson2}
\end{eqnarray}
with the effective one-particle excitation energy
\begin{eqnarray}
E_{k,M\pm} \equiv \sqrt{{\bm p}^2+R_k({\bm p}) +(m_{M\pm}^{\rm eff})^2  }\ . \label{EkM}
\end{eqnarray}
Here we have defined effective masses for the mesons by
\begin{eqnarray}
(m_{M\pm}^{\rm eff})^2 = 4(m_M^2\pm a_M)\ . 
\end{eqnarray}
The propagator in the baryonic sector, i.e., ${\cal S}^4$ - ${\cal S}^5$ and ${\cal P}^4$ - ${\cal P}^5$, is derived by inverting the respective $2\times2$ matrices. A straightforward computation may yield a result with which the Matsubara frequencies $\omega_n$ inhabit numerators as well as denominators. This expression would, however, generate rather complicated expressions after performing the Matsubara summation and is not convenient. Thus, instead, we adopt the following expressions:
\begin{eqnarray}
{\cal D}_{0,k}^{{\cal S}^4-{\cal S}^5} &=& -\frac{1}{2E_{k,B-}}\Bigg[\frac{\lambda}{i\omega_n-\epsilon^{\rm p}_{k,B-}} - \frac{\lambda^T}{i\omega_n+\epsilon^{\rm p}_{k,B-}} \nonumber\\
&& + \frac{\lambda^T}{i\omega_n-\epsilon^{\rm a}_{k,B-}} -  \frac{\lambda^\dagger}{i\omega_n+\epsilon^{\rm a}_{k,B-}} \Bigg] \ , \label{DBaryon1}
\end{eqnarray}
and
\begin{eqnarray}
{\cal D}_{0,k}^{{\cal P}^4-{\cal P}^5} &=& -\frac{1}{2E_{k,B+}}\Bigg[\frac{\lambda}{i\omega_n-\epsilon^{\rm p}_{k,B+}} - \frac{\lambda^T}{i\omega_n+\epsilon^{\rm p}_{k,B+}} \nonumber\\
&& + \frac{\lambda^T}{i\omega_n-\epsilon^{\rm a}_{k,B+}} -  \frac{\lambda^\dagger}{i\omega_n+\epsilon^{\rm a}_{k,B+}} \Bigg] \ ,  \label{DBaryon2}
\end{eqnarray}
where we have introduced a $2\times2$ matrix
\begin{eqnarray}
\lambda= \frac{1}{2}\left(
\begin{array}{cc}
1 & i \\
-i & 1 \\
\end{array}
\right)\ ,
\end{eqnarray}
and
\begin{eqnarray}
\epsilon^{\rm p}_{k,B\pm} = E_{k,B\pm}-2\mu_q\ , \ \ \epsilon^{\rm a}_{k,B\pm} = E_{k,B\pm} + 2\mu_q\ ,
\end{eqnarray}
with the baryonic dispersion relations
\begin{eqnarray}
E_{k,B\pm} \equiv \sqrt{{\bm p}^2+R_k({\bm p}) + (m_{B\pm}^{\rm eff})^2}\ .  \label{EkB}
\end{eqnarray}
Here we have defined the effective masses
\begin{eqnarray}
(m_{B\pm}^{\rm eff})^2 = 8(m_M^2\pm a_M)\ .
\end{eqnarray}
Utilizing the above propagators, we only encounter the following Matsubara summations:
\begin{eqnarray}
&& T\sum_n\frac{1}{(i\omega_n-\epsilon_1)(i\omega_n-\epsilon_2)} = \frac{f_B(\epsilon_2)-f_B(\epsilon_1)}{\epsilon_1-\epsilon_2}\ , \nonumber\\
&& T\sum_n\frac{1}{(i\omega_n-\epsilon)^2}= -\frac{df_B(\epsilon)}{d\epsilon}\ , \label{MatsubaraF}
\end{eqnarray}
with the Bose-Einstein distribution function $f_B(\epsilon) = 1/({\rm e}^{\epsilon/T}-1)$, which are much easier to handle.

\begin{table}[t]
\begin{center}
  \begin{tabular}{c|cccccc}  \hline\hline
 & $\tilde{m}_\Lambda^2$ [GeV$^2$] & $\tilde{a}_\Lambda$ [GeV] & $\tilde{\lambda}_{1,\Lambda}$ & $\tilde{\lambda}_{2,\Lambda}$ & $\tilde{c}_{1,\Lambda}$ & $\tilde{c}_{2,\Lambda}$ \\ \hline
Set (I) & $-0.0426$ & $-0.032$ & $1.25$ & $5$ & $0$ & $0$ \\
Set (II) & $0.385$ & $-0.428$ & $1.25$ & $5$ & $0$ & $0$ \\
\hline \hline
 \end{tabular}
\caption{Two sets of initial parameters at the UV scale $\Lambda=1$ GeV taken from Ref.~\cite{Fejos:2025oxi}. }
\label{tab:Input}
\end{center}
\end{table}

\begin{figure*}[t]
\centering
\hspace*{-0.2cm} 
\includegraphics*[scale=0.52]{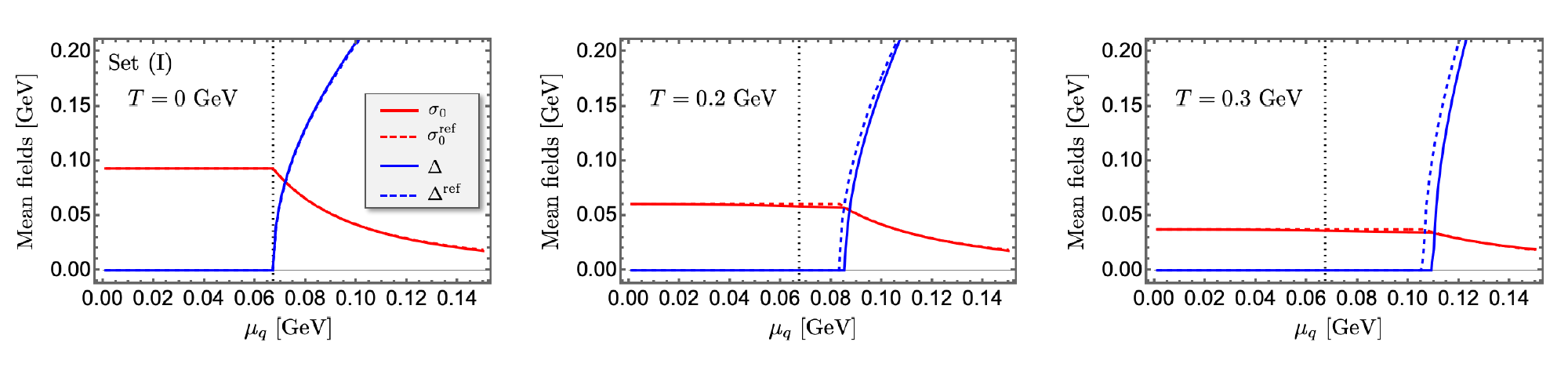}
\caption{$\mu_q$ dependence of the mean fields $\sigma_0$ and $\Delta$ at $T=0,0.2,0.4$ GeV, with the parameter set (I). The dotted vertical line corresponds to $\mu_q = \mathring{M}_\pi/2 = 0.067$ GeV.}
\label{fig:MeanFieldSet1}
\end{figure*}
\begin{figure*}[t]
\centering
\hspace*{-0.2cm} 
\includegraphics*[scale=0.52]{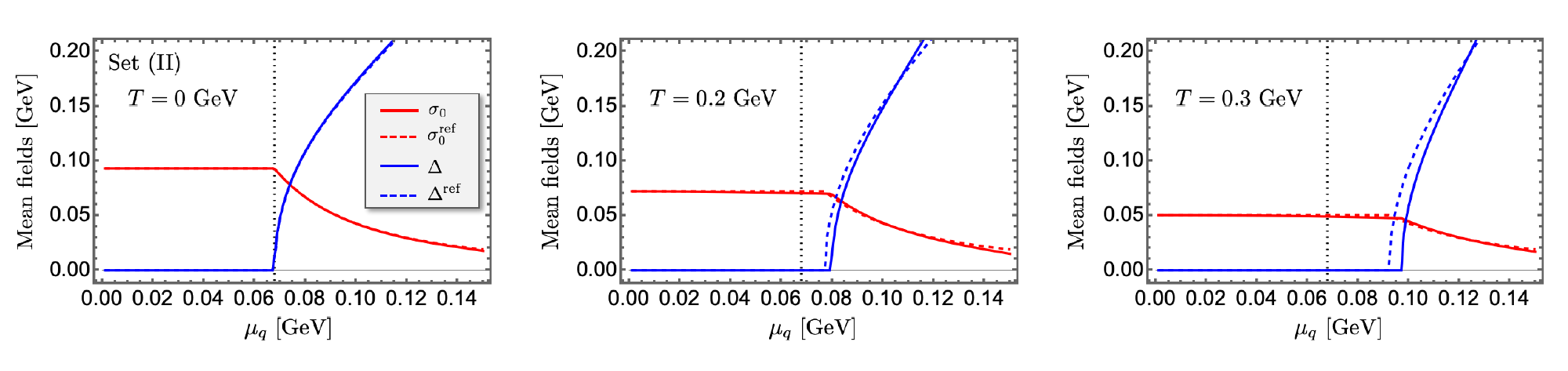}
\caption{$\mu_q$ dependence of the mean fields $\sigma_0$ and $\Delta$ with the parameter set (II). The dotted vertical line corresponds to $\mu_q = \mathring{M}_\pi/2 = 0.068$ GeV.}
\label{fig:MeanFieldSet2}
\end{figure*}

With the help of the above expressions for $D_{0,k}$, the r.h.s. of Eq.~(\ref{RHSExpand}) can be evaluated. The first term in the r.h.s. corresponds to the vacuum contribution, which does not contain any fields $\phi$, and thus can be dropped. The second and third terms will be translated into the r.h.s. of flow equations for $m_{M}^2$, $m_{B}^2$, $\cdots$, by reading off the coefficients of each term, ${\rm tr}[\Sigma_M^\dagger\Sigma_M]$, ${\rm tr}[\Sigma_{B_R}^\dagger\Sigma_{B_R} + \Sigma_{B_L}^\dagger\Sigma_{B_L}]$, $\cdots$. Derivation of all the flow equations is rather complicated and the resultant r.h.s. are too lengthy, therefore, we do not show them here explicitly. In Appendix~\ref{sec:FlowEquation} we present some of the technical details by showing concrete examples.

\section{Numerical results}
\label{sec:Numerical}

\subsection{Notes in solving flow equations}
\label{sec:NoteFlow}

\begin{figure}[t]
\centering
\hspace*{-0.2cm} 
\includegraphics*[scale=0.6]{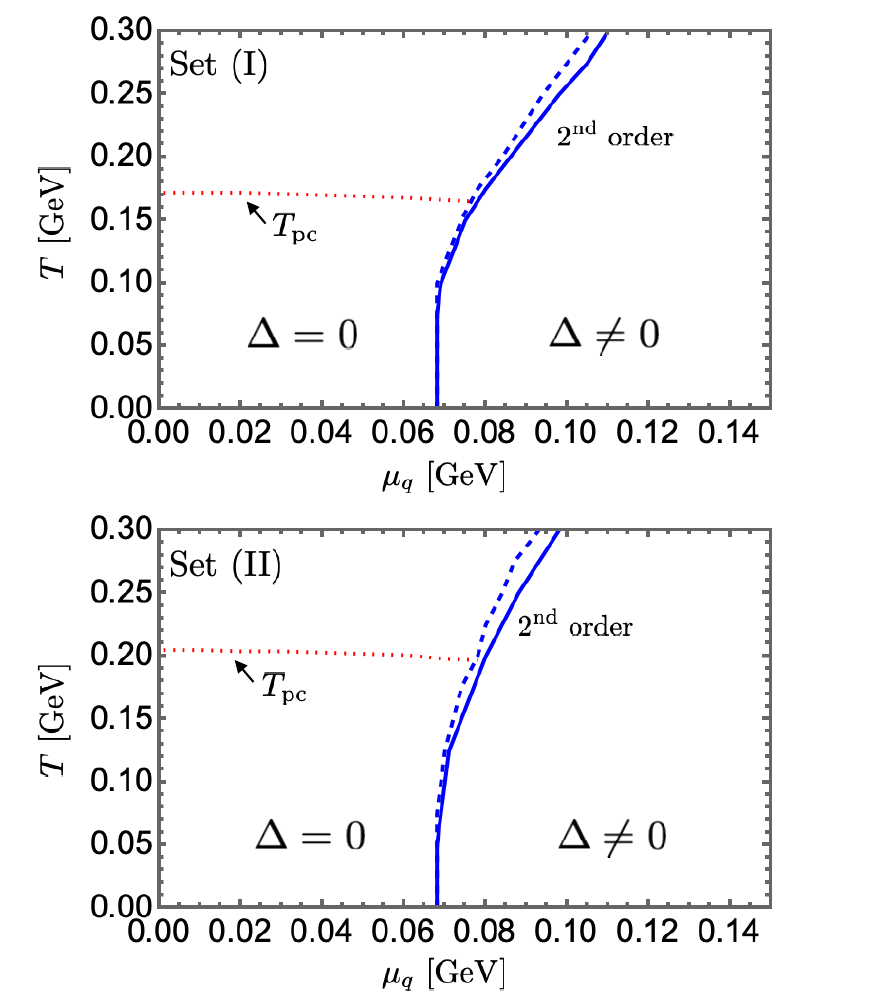}
\caption{PB separating the hadronic ($\Delta=0$) and baryon superfluid ($\Delta\neq0$) phases, for the parameter set (I) (top) and set (II) (bottom). The dashed curve represents the PB evaluated with the $\mu_q$-independent coefficients at each temperature: $m_M^2|_{\mu_q=0}$, $m_B^2|_{\mu_q=0}$, $\cdots$. The dotted curve denotes the pseudocritical temperature ${T}_{\rm pc}$ of the chiral restoration.}
\label{fig:PhaseDiagram}
\end{figure}

\begin{figure*}[t]
\centering
\hspace*{-0.2cm} 
\includegraphics*[scale=0.58]{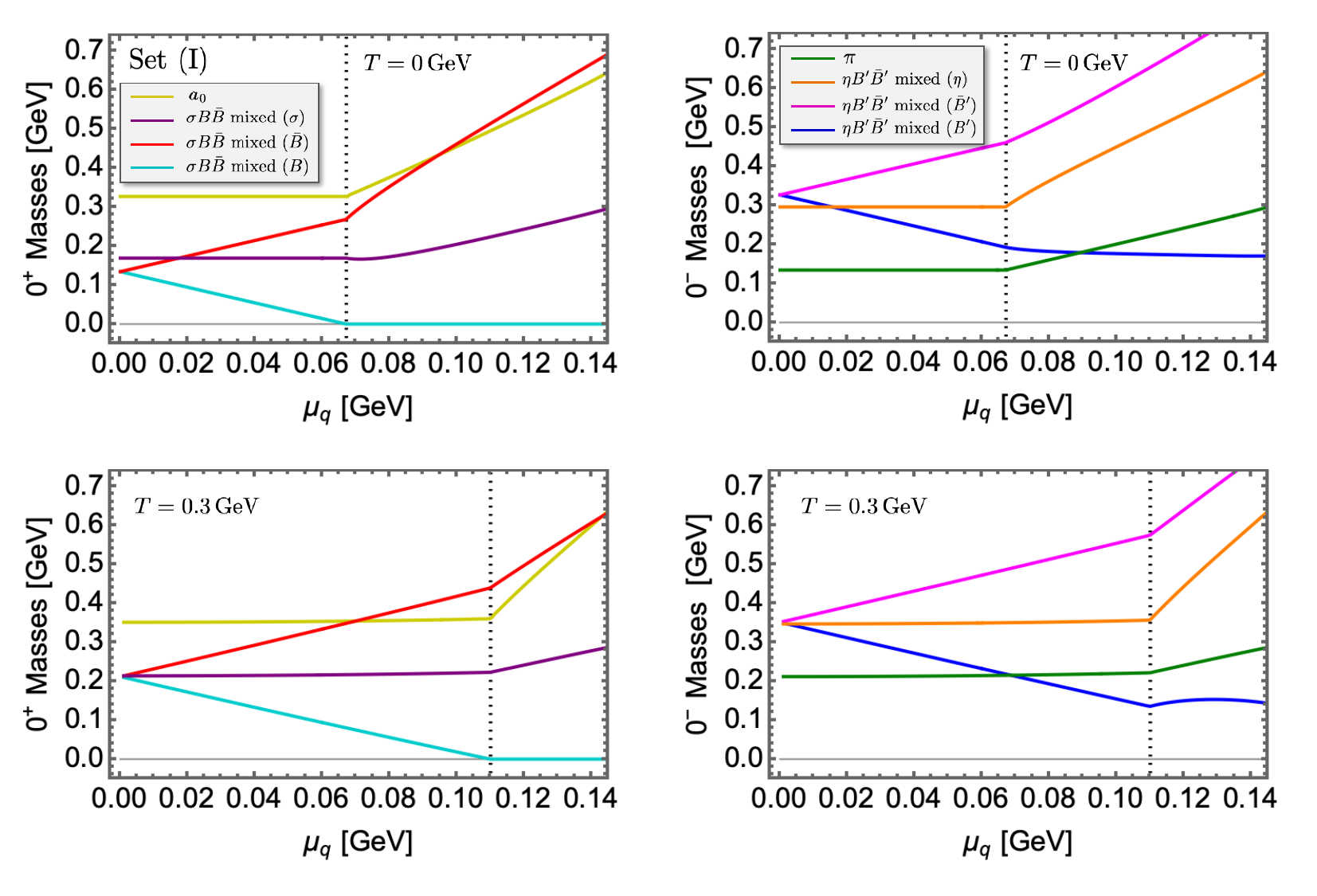}
\caption{$\mu_q$ dependence of the $0^+$ (left) and $0^-$ (right) hadron mass spectrum at $T=0,0.3$ GeV, with the parameter set (I). The vertical line stands for the critical chemical potential separating the hadronic and superfluid phases at each temperature.}
\label{fig:HadronMass1}
\end{figure*}
\begin{figure*}[t]
\centering
\hspace*{-0.2cm} 
\includegraphics*[scale=0.58]{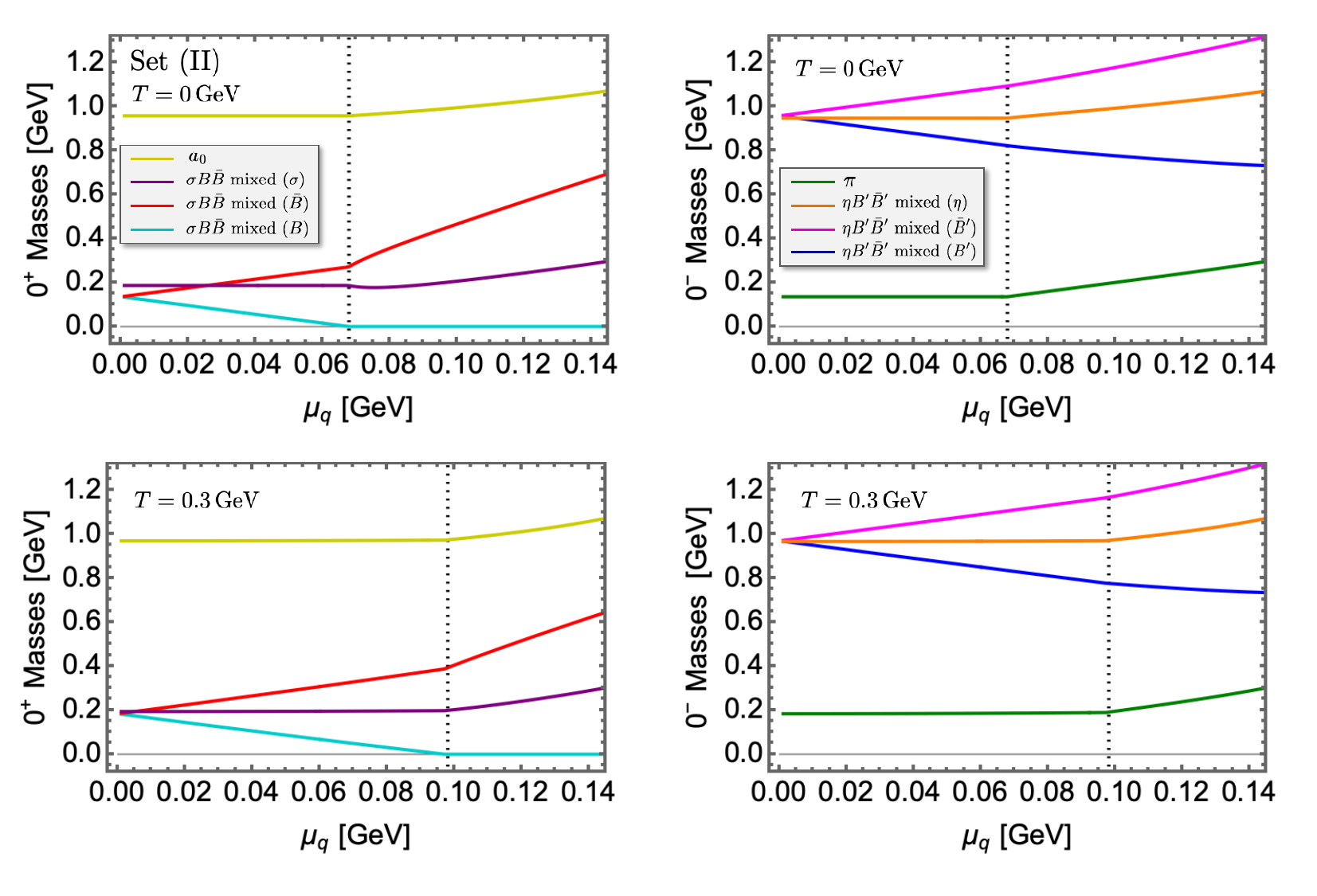}
\caption{$\mu_q$ dependence of the hadron mass spectrum with the parameter set (II).}
\label{fig:HadronMass2}
\end{figure*}

\begin{figure}[t]
\centering
\hspace*{-0.2cm} 
\includegraphics*[scale=0.6]{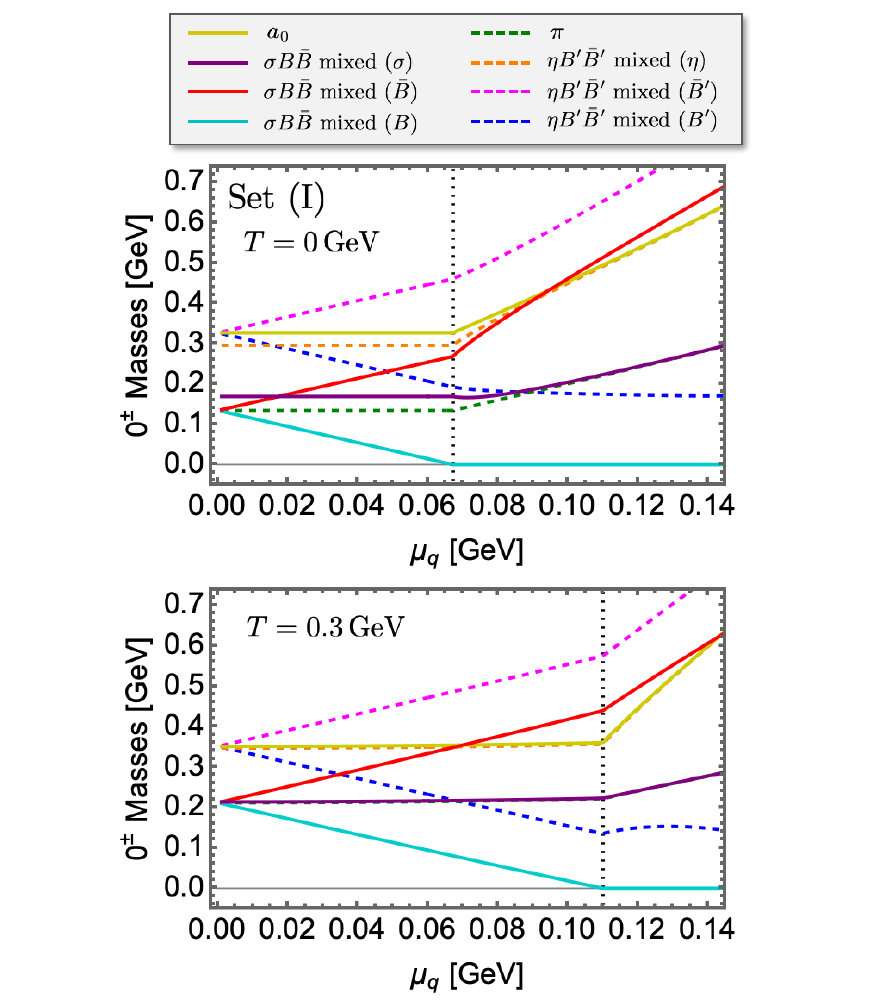}
\caption{$\mu_q$ dependence of the $0^\pm$ hadron masses at $T=0$ GeV (top) and $T=0.3$ GeV (bottom), with the parameter set (I).}
\label{fig:ChiralPartner}
\end{figure}

\begin{figure}[t]
\centering
\hspace*{-0.2cm} 
\includegraphics*[scale=0.6]{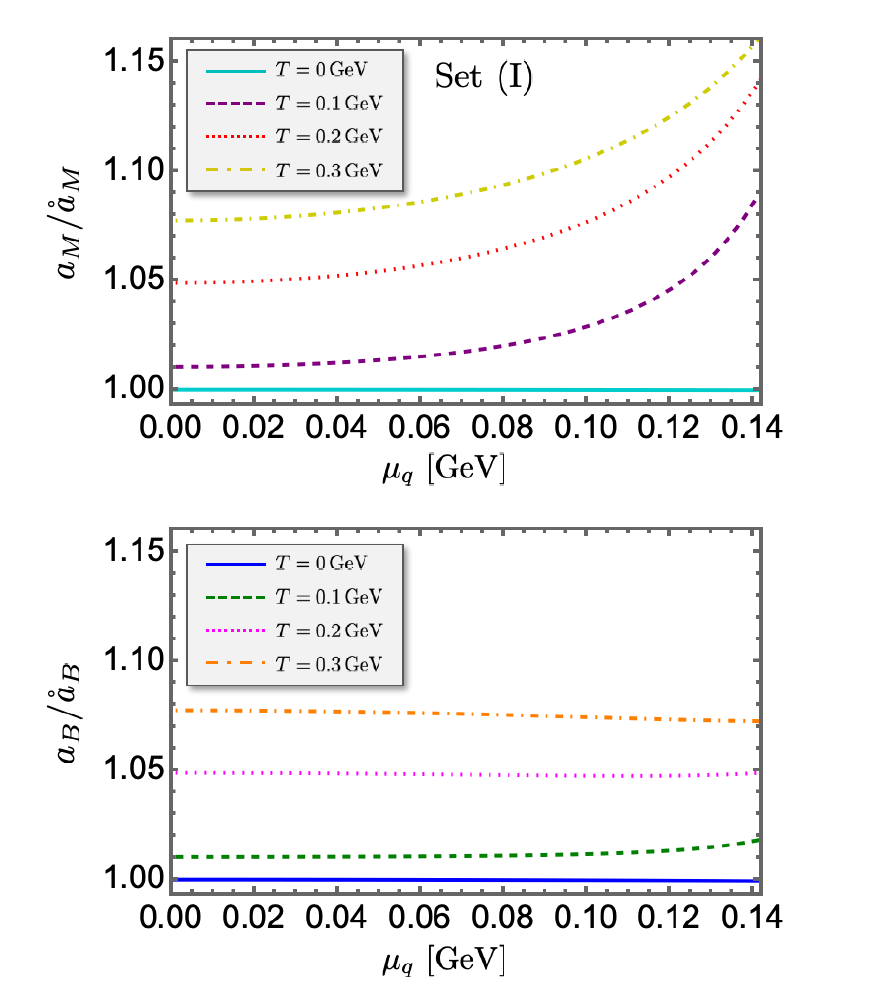}
\caption{$\mu_q$ dependence of the mass-dimension $+2$ anomaly couplings $a_M$ (top) and $a_B$ (bottom) at $T=0,0.1,0.2,0.3$ GeV, with the parameter set (I).}
\label{fig:Anomaly1}
\end{figure}

\begin{figure}[t]
\centering
\hspace*{-0.2cm} 
\includegraphics*[scale=0.6]{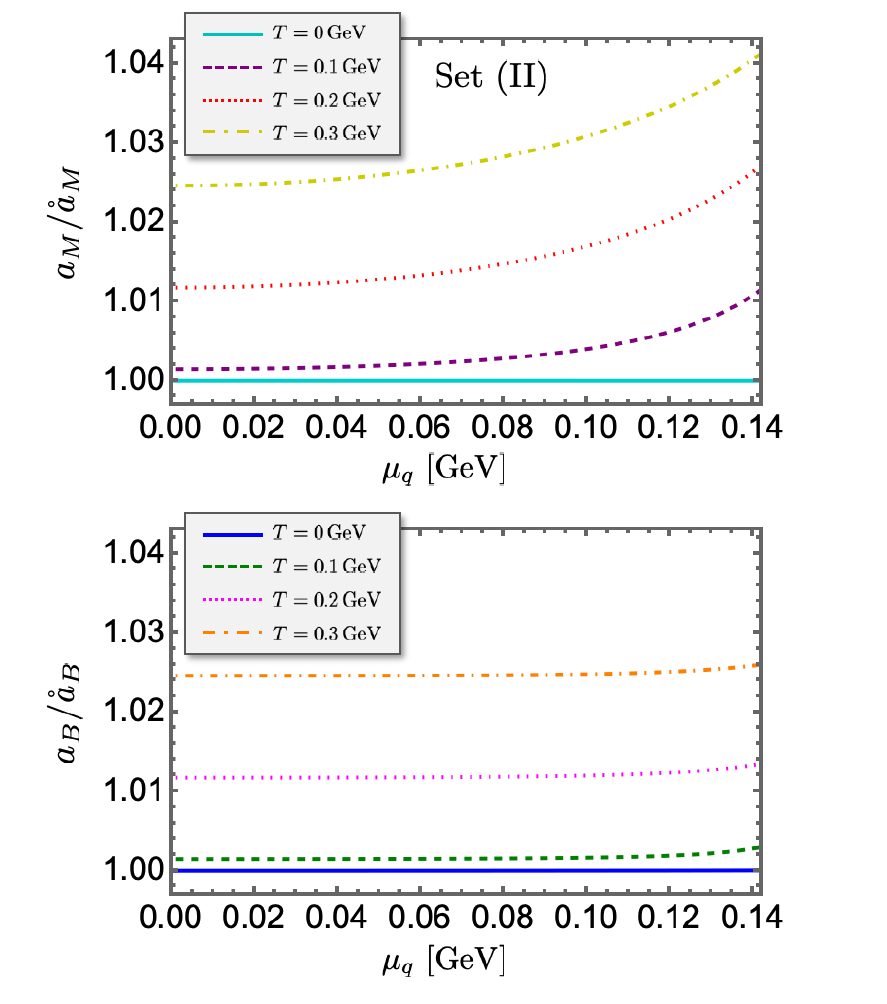}
\caption{$\mu_q$ dependence of the anomaly couplings $a_M$ and $a_B$ with the parameter set (II).}
\label{fig:Anomaly2}
\end{figure}

Before proceeding with the numerical analysis, we explain important details on solving the flow equations.

First, we present our criterion in determining initial conditions for the $17$ couplings. 
Although there is no a priori knowledge of their respective values at the initial scale, it is reasonable to assume the Pauli-G\"{u}rsey $SU(4)$ symmetry in the UV, which is also done in perturbative analyses~\cite{Suenaga:2025sln}, since the potential at the UV scale essentially corresponds to a classical theory. This assumption leads to the following relationships between the couplings:
\begin{eqnarray}
&& m_{M,\Lambda}^2=2m_{B,\Lambda}^2= \frac{\tilde{m}^2_{\Lambda}}{4}\ , \nonumber\\
&& \lambda_{M1,\Lambda} = 4 \lambda_{B1,\Lambda} = \gamma_{1,\Lambda}+\frac{\gamma_{2,\Lambda}}{2}  = \frac{\tilde{\lambda}_{1,\Lambda}}{16}\ ,  \nonumber\\
&& \lambda_{M2,\Lambda} = 8\lambda_{B2,\Lambda} =-\gamma_{2,\Lambda} = \frac{\tilde{\lambda}_{2,\Lambda}}{32} \ , \nonumber\\
&& a_{M,\Lambda}=2a_{B,\Lambda} = \frac{\tilde{a}_\Lambda}{4} \ , \ \ c_{M1,\Lambda} = 4c_{B1,\Lambda} = d_{3,\Lambda} = \frac{\tilde{c}_{1,\Lambda}}{16} \ , \nonumber\\
&&  c_{M2,\Lambda} = 4c_{B2,\Lambda} = 2d_{1,\Lambda}=2d_{2,\Lambda}   = \frac{\tilde{c}_{2,\Lambda}}{16} \ ,
\end{eqnarray}
as shown in Eq.~(\ref{SU4LimitParameter}), where $\tilde{m}^2_{\Lambda}$, $\tilde{\lambda}_{1,\Lambda}$, $\cdots$ represent the corresponding $SU(4)$ symmetric parameters (for details see Appendix~\ref{sec:PotentialSU4}). Furthermore, as done in Ref.~\cite{Fejos:2025oxi}, we make use of the large-$N_c$ argument to reduce the number of unknown initial parameters~\cite{Witten:1979kh}. This assumption claims 
\begin{eqnarray}
\tilde{\lambda}_{2,\Lambda}=4\tilde{\lambda}_{1,\Lambda}\ , \ \ \tilde{c}_{1,\Lambda} = \tilde{c}_{2,\Lambda} = 0\ ,
\end{eqnarray}
at the UV scale. 

The above reductions leave us with only three parameters to be determined: $\tilde{m}_\Lambda^2$, $\tilde{\lambda}_{1,\Lambda}$ and $\tilde{a}_\Lambda$, which is the same as in our previous study at finite temperature with $\mu_q=0$~\cite{Fejos:2025oxi}. Hence, in the present work we employ the same initial parameter sets. We tabulate those values in Table~\ref{tab:Input}. These initial parameters yield $\mathring{M}_\pi = 0.14$ GeV and $\mathring{f}_\pi = \mathring{\sigma}_0 = 0.093$ GeV in the vacuum, i.e., at $T=\mu_q=0$.\footnote{The symbol ``$\mathring{X}$'' stands for the corresponding vacuum value.} In addition, the parameter set (I) and set (II) in Table~\ref{tab:Input} lead to the vacuum masses 
$\mathring{M}_\eta=0.3$ GeV and $\mathring{M}_\eta=0.95$ GeV, when the IR limit $k\to0$ is taken.\footnote{As explained below, in the present work we do not take the exact IR limit but employ a finite IR cutoff of $k_{\rm IR} = 0.3$ GeV. Thus, the vacuum mass values are slightly changed, but the deviations are only a few \%.} That is, the parameter set (I) and set (II) correspond to small and large $U(1)_A$ anomaly effects, respectively. We note that the UV scale is chosen to be $\Lambda=1$ GeV.

The next important thing to note is related to the IR limit. The excitations contributing to the r.h.s. of flow equations are bosons with the dispersion relation~(\ref{EkM}) or~(\ref{EkB}). After the ${\bm p}$ integration, the dispersion relations will be replaced by $\sqrt{k^2+(m_{M\pm}^{\rm eff})^2}$ and $\sqrt{k^2+(m_{B\pm}^{\rm eff})^2}$, as shown in Appendix~\ref{sec:FlowEquation}.  At adequately large $\mu_q$, the baryonic dispersion $\sqrt{k^2+(m_{B\pm}^{\rm eff})^2}$ gets close to $2\mu_q$: $\sqrt{k^2+(m_{B\pm}^{\rm eff})^2}\sim 2\mu_q$, during the flows. In such a case, the occupation probability, $f_B(\sqrt{k^2+(m_{B\pm}^{\rm eff})^2}-2\mu_q)$, is enhanced due to the attractive boson statistics at low energy. Accordingly, the r.h.s of the flow equations blow up at some small $k$, which obviously shows the breakdown of the present approximation~\cite{Terazaki:2024evv}. Hence, in order to avoid this artifact of the approximation, we need to limit the lower boundary of the $k$ integration to some nonzero IR scale: $k=k_{\rm IR}$. In what follows, we universally adopt $k_{\rm IR}=0.3$ GeV for any $T$ and $\mu_q$, therefore, in all of the QC$_2$D medium our exploration is done in a consistent fashion. The $k_{\rm IR}$ value yields a reasonable convergence for each flow equation and the resulting couplings do not change much with respect to $k$ at this point already. A detailed argument is provided in Appendix~\ref{sec:KIRDetermination}.



The final issue we address here is the Silver-Blaze property. Within perturbative analyses based on chiral models, the Silver-Blaze property, meaning that all observables (except for the diquark baryon masses) are independent of $\mu_q$, is derived in the hadronic phase at zero temperature~\cite{Kogut:2000ek,Ratti:2004ra,Suenaga:2022uqn}. Our present FRG analysis does not exactly yield this property at $T=0$, but the violation is found to be very tiny; the Silver-Blaze property at $T=0$ holds with high accuracy in our framework. We wish to note that the Silver-Blaze property is indeed supported by lattice simulations, particularly at sufficiently low $T$~\cite{Iida:2024irv}.

\subsection{Phase diagram}
\label{sec:MeanFields}

The phase structure of QC$_2$D medium is quantified by its symmetry properties such as chiral-symmetry restoration and $U(1)_B$-symmetry breaking. First, we focus on the $\mu_q$ dependency of the mean fields
\begin{eqnarray}
\sigma_0 \equiv \langle {\cal S}^0\rangle \ , \ \ \Delta \equiv \langle{\cal P}^5\rangle\ ,
\end{eqnarray}
which are regarded as order parameters of chiral-symmetry and $U(1)_B$-symmetry breakings, respectively. These mean-field values are evaluated by solving the respective gap equations: $\partial V/\partial\sigma_0=0$ and $\partial V/\partial\Delta=0$ at the IR scale.

Depicted in Figs.~\ref{fig:MeanFieldSet1} and~\ref{fig:MeanFieldSet2} are the resultant $\mu_q$ dependencies of ${\sigma}_0$ and $\Delta$ at $T=0,0.2,0.4$ GeV, with the parameter set (I) and set (II) defined in Table~\ref{tab:Input}, respectively. The vertical dotted lines in the figures denote $\mu_q = \mathring{M}_\pi/2$, where the vacuum pion masses, evaluated at $k_{\rm IR} = 0.3$ GeV, are $\mathring{M}_\pi = 0.134$ GeV and $\mathring{M}_\pi=0.136$ GeV for the sets (I) and (II), respectively. The dashed curves ${\sigma}^{\rm ref}_0$ and ${\Delta}^{\rm ref}$ are drawn by solving the gap equations with the couplings $m_M^2$, $m_B^2$, $\cdots$ evaluated at their $\mu_q=0$ values, $m_M^2|_{\mu_q=0}$, $m_B^2|_{\mu_q=0}$, $\cdots$, at each temperature. The dashed curves, therefore, show the importance of the implicit $\mu_q$ dependencies of the couplings, obtained from the flow equations at fixed $T$.

The left-most panels in Figs.~\ref{fig:MeanFieldSet1} and~\ref{fig:MeanFieldSet2} indicate that at $T=0$, the diquark gap $\Delta$ is always zero up to $\mu_q = \mathring{M}_\pi/2$, and above this critical point it acquires a nonzero value. That is, $\mu_q = \mathring{M}_\pi/2$ is the critical chemical potential separating the hadronic phase and baryon superfluid phase at vanishing temperature\footnote{This critical chemical potential is numerically confirmed with good accuracy.}, which is consistent with lattice results~\cite{Iida:2024irv}. As the temperature increases, the phase-transition point shifts to the right due to the inhibition caused by hadronic thermal excitations. Meanwhile, the chiral condensate, $\sigma_0$, is almost constant in the hadronic phase, and it monotonically decreases in the superfluid phase and show chiral restoration.  It should be noted that the phase transition to the superfluid phase is always of second order.

Comparing the solid and dashed curves, one can see that the $\mu_q$-dependent RG flows slightly delay the phase transition. At lower temperatures, this effect becomes milder. In particular, at $T=0$ the solid and dashed curves almost overlap. In this case, indeed, we have numerically confirmed that none of the couplings are sizably modified at finite $\mu_q$. This stable property implies manifestation of the Pauli-G\"{u}rsey $SU(4)$ symmetry with high accuracy even at finite $\mu_q$ as long as $T=0$, although in general the symmetry is broken into $SU(2)_L\times SU(2)_R\times U(1)_B$. The property is also reminiscent of the Silver-Blaze property up to $\mu_q=\mathring{M}_\pi/2$.

In order to visualize the phase structure, we depict the phase boundary (PB) separating the hadronic and superfluid phases with the two parameter sets in Fig.~\ref{fig:PhaseDiagram} by blue curves. Again, the dashed curve is drawn by solving the gap equations with dropping the $\mu_q$ dependencies of the couplings, $m_M^2|_{\mu_q=0}$, $m_B^2|_{\mu_q=0}$, $\cdots$, at each temperature. While the phase transition to the superfluid phase is always of second-order, the chiral-symmetry restoration to the hadronic phase is a crossover. The red dotted curve represents the pseudocritical temperature $T_{\rm pc}$ of this crossover transition defined at which $\partial^2\sigma_0/\partial T^2=0$ is satisfied, at some fixed $\mu_q$. After the red curve hits the PB of superfluidity, the crossover curve aligns with the solid blue curve.

Figure~\ref{fig:PhaseDiagram} indicates that the chiral pseudocritical temperature $T_{\rm pc}$ is less sensitive to the chemical potential in the hadronic phase. As for the transition to the superfluid phase, its critical chemical potential is given by $\mu_q\approx \mathring{M}_\pi/2$, independently of $T$, as long as the latter is not too large. This behavior is consistent with the lattice~\cite{Boz:2013rca,Buividovich:2020dks}. At higher $T$, our results indicate that the critical chemical potential depends mildly on $T$, while the lattice data do not; the lattice results seem to imply an almost constant critical temperature at any $\mu_q$. 

The above discrepancy could be understood, within the present FRG framework, by the breakdown of our approximation at the IR. As examined in Fig.~\ref{fig:KDepSet2} in Appendix~\ref{sec:KIRDetermination}, at larger $\mu_q$ and $T$ the convergence of the couplings is getting worse, meaning that the truncation at $k_{\rm IR} = 0.3$ GeV underestimates the flow results. Or, physically speaking, this bad behavior at the IR implies the necessity of including other fundamental fluctuations, such as explicit quark degrees of freedom at large $\mu_q$ and $T$, as naturally inferred from deconfinement. In fact, other approaches incorporating quarks, such as the Nambu--Jona-Lasinio model or the quark-meson-diquark model result in larger $T$ dependencies of the PB at high $\mu_q$ even at the mean-field level~\cite{Ratti:2004ra,Sun:2007fc,Strodthoff:2011tz,Strodthoff:2013cua}.

\subsection{Hadron mass spectrum}
\label{sec:HadronMass}

Next, we present the mass spectrum of the low-lying $0^\pm$ hadrons in the QC$_2$D medium, being significant observables on the lattice to probe the chiral restoration and the onset of baryon superfluidity at finite $T$ and $\mu_q$. The mass formulas are supplemented in Appendix~\ref{sec:HadronMassApp}.

Depicted in Figs.~\ref{fig:HadronMass1} and~\ref{fig:HadronMass2} are the resultant $\mu_q$ dependencies of the mass spectrum at $T=0,0.3$ GeV, with the parameter set (I) and set (II), respectively. The vertical dotted line stands for the respective critical chemical potential separating the hadronic and superfluid phases. In the superfluid phase, $\sigma$, $B$, and $\bar{B}$ mix, as do $\eta$, $B'$ and $\bar{B}'$, due to the violation of $U(1)_B$ symmetry, also indicated in the insets. The hadrons denoted in parenthesis stand for the pure state with no mixings in the hadronic phase. At $\mu_q=0$, one can see $M_{a_0}=M_{B'}=M_{\bar{B}'}$ and $M_{\pi}=M_B=M_{B'}$, arising from the Pauli-G\"{u}rsey $SU(4)$ symmetry at any temperature. The figures also imply $M_\eta \approx M_{a_0,B',\bar{B}'}$, while $M_{\sigma}\approx M_{\pi,B,\bar{B}}$ at $\mu_q=0$, although there are no corresponding symmetry relations. These rather small mass differences are due to the suppression of the four-point couplings, which is required to obtain the crossover chiral restoration at finite $T$ with small $\mu_q$ consistent with the lattice~\cite{Fejos:2025oxi}.

Within the present FRG framework, in the $\sigma B\bar{B}$-mixed sector we see a massless mode emerging in the superfluid phase, which is regarded as a Nambu-Goldstone (NG) boson associated with the spontaneous breakdown of $U(1)_B$ symmetry at any temperature, similarly to other works based on perturbative treatments~\cite{Kogut:1999iv,Ratti:2004ra,Suenaga:2025sln}. The pion mass in the superfluid phase is given by $M_\pi\approx 2\mu_q$, in particular, at $T=0$, $M_\pi= 2\mu_q$ holds with high accuracy.\footnote{The linear $\mu_q$ dependencies of the mass of massive NG mode in cold medium were discussed in Refs.~\cite{Nicolis:2012vf,Watanabe:2013uya} model-independently.} 

With the parameter set (I), $M_\eta$, and accordingly $M_{a_0,B',\bar{B}'}$ do not become large, reflecting less significant $U(1)_A$ anomaly effects. Hence, the $B'$ mass, i.e., the blue curve, decreases monotonically in the hadronic phase and gets smaller than the pion mass at some $\mu_q$, as indicated in Fig.~\ref{fig:HadronMass1}. This mass inversion is one of the remarkable features of the hadron mass spectrum in the superfluid phase. At $T=0$, the lowest $\eta B'\bar{B}'$ mode in the superfluid phase monotonically decreases, drawing a gentle convex-downward curve with both parameter sets, which is consistent with the previous work~\cite{Suenaga:2022uqn}. At $T=0.3$ GeV, however, the mode exhibits a convex-upward dependency on $\mu_q$ with the parameter set (I), whereas with the set (II) it does not. The former peculiar behavior might have been obtained as the lowest $\eta B'\bar{B}'$-mixed mode already gets significantly light at the critical chemical potential, $\mu_q\sim 0.11$ GeV.

At finite $T$ and $\mu_q$, chiral symmetry is partially restored as implied by the mean field plots in Figs.~\ref{fig:MeanFieldSet1} and~\ref{fig:MeanFieldSet2}. This restoration leads to mass degeneracies of the chiral-partner hadrons, as broadly examined in both QC$_2$D and ordinary three-color QCD~\cite{Hatsuda:1994pi,Rapp:1999ej,Suenaga:2025sln}. In order to visualize the mass degeneracies, in Fig.~\ref{fig:ChiralPartner} we depicted the mass spectrum of all the $0^\pm$ hadrons using the parameter set (I). We have plotted the mass spectrum at $T=0$ and $T=0.3$ GeV to take a closer look at temperature effects on the chiral restoration. The figure demonstrates that at sufficiently large $\mu_q$, the masses of $(\sigma,\pi)$ and $(a_0,\eta)$ tend to degenerate. As for the remaining hadrons, one might expect mass degeneracy in $(B,B')$ and $(\bar{B},\bar{B}')$, but these excitations are not related by any chiral transformation and do not necessarily degenerate even when chiral symmetry is sufficiently restored. These mass degeneracies seem to take place only if the $U(1)_A$ anomaly is significantly suppressed~\cite{Suenaga:2022uqn}.

Comparing the top and bottom panels in Fig.~\ref{fig:ChiralPartner}, one can see that the mass degeneracies between the parity partners, particularly $(\sigma,\pi)$ and $(a_0,\eta)$, are well realized at $T=0.3$ GeV even in the hadronic phase. These smaller mass differences reflect the partial chiral restoration at finite temperature. We note that $T=0.3$ GeV lies above the pseudocritical temperature $T_{\rm pc}$, where the chiral condensate $\sigma_0$ is already sufficiently reduced: $\sigma_0/\mathring{\sigma}_0|_{T=0.3\, {\rm GeV}} \approx 0.4$, with the parameter set (I).

We note that the $U(1)_A$ anomaly effect does not directly have an influence on the chiral-partner structure. Thus, when taking the parameter set (II), which incorporates a comparably large $U(1)_A$ anomaly effect, we obtain the mass degeneracies of ($\sigma,\pi$) and $(a_0,\eta)$ in a similar fashion as above.

\subsection{Anomaly coefficients $a_M$ and $a_B$ in medium}
\label{sec:Anomaly}

In Ref.~\cite{Fejos:2025oxi}, it was argued that the $U(1)_A$ anomaly couplings are enhanced at finite temperature due to hadronic fluctuations, before they vanish at the high-$T$ limit owing to the Debye screening of electric gluons~\cite{Gross:1980br,Rapp:1999qa}. In order to shed light on the fate of $U(1)_A$ anomaly in the dense regime of QC$_2$D, in this subsection we examine the $\mu_q$ dependence of the anomaly coefficients.

Our generalized LSM has nine anomaly coefficients, as presented in Eq.~(\ref{VAGeneral1}). Among them, $a_M$ and $a_B$ have mass-dimension $+2$, while the remaining ones have mass-dimension $0$. Hence, here we only concentrate on the relevant $a_M$ and $a_B$ as representatives. Depicted in Figs.~\ref{fig:Anomaly1} and~\ref{fig:Anomaly2} are the resultant $\mu_q$ dependence of $a_M$ and $a_B$ at $T=0,0.1,0.2,0.3$ GeV, with the parameter set (I) and (II). We have normalized the coefficients by their respective vacuum values to see modifications in the medium in more detail. The figures indicate that $a_M$ at finite temperature exhibits a substantial enhancement with increasing $\mu_q$. In particular, using the parameter set (I), with which the $U(1)_A$ anomaly effect in the vacuum is less estimated, yields a more prominent enhancement; one can see the $\sim 15\%$ change in Fig.~\ref{fig:Anomaly1} at $T=0.3$ GeV and $\mu_q\sim 2\mathring{M}_\pi\sim0.14$ GeV, while the parameter set (II) in Fig.~\ref{fig:Anomaly2} exhibits only a $\sim4\%$ enhancement.
 
 Meanwhile, the $\mu_q$ dependence of $a_B$ is not as significant in case of both parameter sets at any $T$, although the enhancement with respect to $T$ is still manifest. Since $a_M$ and $a_B$ are anomaly coefficients for mesons and baryons, the obtained results imply that the mesonic part of the anomaly is largely modified at finite density, while the baryonic one is less influenced by a dense medium. Although these clear differences are expected to provide useful information on the fate of the anomaly in a cold and dense medium, the mechanism behind the obtained structure remains to be unveiled. It should be noted that the truncation of the flows at $k_{\rm IR} = 0.3$ GeV might underestimate the values of the coefficients, particularly at higher temperatures, as showed by $m_M^2$ in Appendix~\ref{sec:KIRDetermination}, as an example. Therefore, $a_M$ at finite $T$ and $\mu_q$, presumably, is further enhanced.
  

Figures~\ref{fig:Anomaly1} and~\ref{fig:Anomaly2} also indicate that neither $a_M$, nor $a_B$ shows a substantial enhancement at zero temperature, regardless of which parameter set was used. As explained in Sec.~\ref{sec:MeanFields}, it has been numerically confirmed that none the couplings are sizeably modified at finite $\mu_q$ as long as we stick to vanishing temperature.

\begin{figure}[t]
\centering
\hspace*{-0.2cm} 
\includegraphics*[scale=0.6]{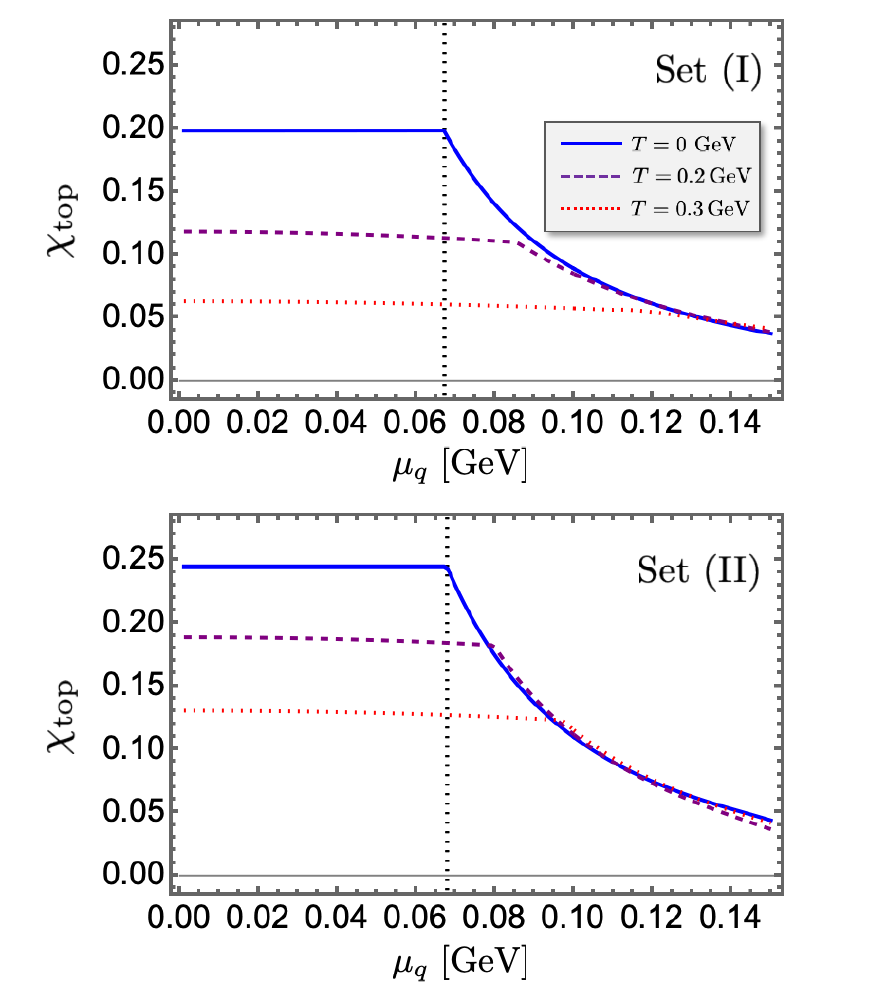}
\caption{$\mu_q$ dependence of the normalized topological susceptibility ${\chi}_{\rm top}$, defined in Eq.~(\ref{NormChiTop}) at $T=0,0.2,0.3$ GeV, with the parameter set (I) (top) and set (II) (bottom). The vertical dotted line corresponds to $\mu_q=\mathring{M}_\pi/2$.}
\label{fig:TopSusceptibility}
\end{figure}

\subsection{Topological susceptibility}
\label{sec:TopSusceptibility}

One of the most important observables in association with the $U(1)_A$ axial anomaly is the topological susceptibility. In this subsection, we present the corresponding numerical results in the QC$_2$D medium.

The topological susceptibility is defined by a two-point function of the gluon topological operator, $Q$, at vanishing momentum: $\bar{\chi}_{\rm top} \equiv i^{-1}\int d^4x\langle Q(x) Q(0)\rangle $, where $Q = (g_s^2/64\pi^2)\epsilon^{\mu\nu\rho\sigma}G_{\mu\nu}^a G_{\rho\sigma}^a$ with the gluon field strength $G_{\mu\nu}^a$ and the strong coupling $g_s$. The $U(1)_A$ anomaly is related to the axial transformation of the quark fields, which implies that the topological susceptibility should be described solely by quark operators. In fact, with the help of the axial Ward-Takahashi identity, as well as Fujikawa's method, one can rewrite the topological susceptibility in the form of the susceptibility functions of fermionic operators $\bar{\psi}i\gamma_5\tau_f^a\psi$ and $\bar{\psi}i\gamma_5\psi$~\cite{GomezNicola:2016ssy,Kawaguchi:2020qvg}. Within the present LSM framework, the resultant $\bar{\chi}_{\rm top}$ reads~\cite{Kawaguchi:2023olk}
\begin{eqnarray}
\bar{\chi}_{\rm top} = i\frac{\mathring{M}_\pi^4\mathring{\sigma}_0^2}{4}\big[D_\pi(0) - D_\eta(0)\big]\ ,
\end{eqnarray}
where $D_\pi(p)$ and $D_\eta(p)$ are two-point functions of pion and $\eta$ meson, respectively. The pion two-point function always take the form of
\begin{eqnarray}
D_\pi(p) = \frac{i}{p^2-M_\pi^2}\ ,
\end{eqnarray}
with the pion mass given in Eq.~(\ref{PiA0Mass}). Note that, the $\eta$ propagator is not straightforwardly derived owing to the mixing with $B'$ and $\bar{B}'$ in the superfluid phase. This two-point function is defined by the ${\cal P}^0{\cal P}^0$ component of the propagator matrix $D_{\eta B'\bar{B}'}$:
\begin{eqnarray}
D_\eta(p) =  [D_{\eta B'\bar{B}'}(p)]_{{\cal P}^0{\cal P}^0}\ ,
\end{eqnarray}
which is obtained by inverting the matrix~(\ref{DEtaBBp}).

Depicted in Fig.~\ref{fig:TopSusceptibility} is the $\mu_q$ dependence of the normalized topological susceptibility
\begin{eqnarray}
\chi_{\rm top} \equiv \frac{\bar{\chi}_{\rm top}}{\mathring{M}_\pi^2\mathring{\sigma}_0^2} =i \frac{\mathring{M}_\pi^2}{4}\big[D_\pi(0) - D_\eta(0)\big]\ , \label{NormChiTop}
\end{eqnarray}
at $T=0,0.2,0.3$ GeV, with both the parameter sets (I) and (II). The vertical dotted line corresponds to the critical chemical potential at $T=0$: $\mu_q=\mathring{M}_\pi/2$. The plots indicate that a stronger $U(1)_A$ anomaly induces a magnified topological susceptibility, as expected. One notes that in the hadronic phase $\chi_{\rm top}$ is comparably more stable against $\mu_q$ at any temperature, which corresponds to the weak $\mu_q$ dependence of $M_\pi$ and $M_\eta$. In the superfluid phase, meanwhile, the susceptibility exhibits a smooth reduction, which simply reflects that the chiral restoration is approximately proportional to $\mu_q^{-2}$, as argued in Ref.~\cite{Kawaguchi:2023olk} in detail. Although the enhancement of the anomaly coefficients were already found in Sec.~\ref{sec:Anomaly}, the chiral restoration in the superfluid phase is more drastic, as explicitly demonstrated in Figs.~\ref{fig:MeanFieldSet1} and~\ref{fig:MeanFieldSet2}. Therefore, when analyzing the topological susceptibility, the actual anomaly enhancement is hidden by the significant reduction of the chiral condensate.

The suppression of $\chi_{\rm top}$ in the superfluid phase seems to be consistent with those lattice results, where the temperature is high enough~\cite{Astrakhantsev:2020tdl}. However, at sufficiently low $T$, recent lattice data indicate an almost constant behavior of $\chi_{\rm top}$ with $\mu_q$, even in the superfluid phase~\cite{Iida:2024irv}. In order to pin down the fate of topological susceptibility in the dense regime, it is inevitable to incorporate the quark-gluon dynamics appropriately into the LSM.

\section{Conclusions}
\label{sec:Conclusions}

In this paper, we have investigated the phase structure, hadron masses, the $U(1)_A$ anomaly effects on hadrons, and the topological susceptibility in QC$_2$D medium. We have employed the linear sigma model with $N_f=2$~\cite{Suenaga:2025sln}, in which fluctuation effects are incorporated through the FRG method. Within the LPA, the effective potential at any flow momentum $k$ is assumed to retain the same form as the classical one, preserving $SU(2)_L \times SU(2)_R$ chiral, $U(1)_B$, parity, and time-reversal symmetries. 

Our results show that the phase transition separating the hadronic and baryon superfluid phases are of second-order at any temperature $T$ and quark chemical potential $\mu_q$. At $T=0$, this phase transition occurs at $\mu_q=\mathring{M}_\pi/2$ (here $\mathring{M}_\pi/2$ is the vacuum pion mass) as the perturbative analysis claims, reflecting the Pauli-G\"{u}rsey $SU(4)$ symmetry with high accuracy. We have also numerically confirmed that none of the couplings are sizably modified at finite $\mu_q$ when $T=0$. As long as $T$ is small, the critical chemical potential is given by $\mu_q\approx\mathring{M}_\pi/2$, independently of $T$. At high $T$, our approach yields a mild $T$ dependence of the critical chemical potential, which is inconsistent with the lattice data, where a strong $T$ dependence is observed.

Within the present FRG analysis, appearance of a massless hadron in the iso-singlet and $0^+$ channel is confirmed at any $T$ in the superfluid phase, which is understood as an Nambu-Goldstone boson associated with the spontaneous breaking of $U(1)_B$ symmetry. Mass degeneracies of chiral partners, i.e., $(\sigma,\pi)$ or $(a_0,\eta)$, are also demonstrated, followed by the restoration of chiral symmetry in medium. 

We have found that the anomaly couplings with mass-dimension of $+2$ are approximately constant against increasing $\mu_q$ at vanishing temperature. However, at finite $T$, these coefficients, particularly the mesonic anomaly coupling, are enhanced with $\mu_q$. Therefore, we can conclude that the effect of the $U(1)_A$ anomaly at finite $T$ and $\mu_q$ is enhanced, as hadronic fluctuations are incorporated into the system. Despite the anomaly enhancement, the topological susceptibility is found to be suppressed as we move into the superfluid phase, following the evaporation of the chiral condensate, which leads to the restoration of chiral symmetry. This shows that even though the anomaly couplings increase with both $T$ and $\mu_q$, the topological susceptibility does not, therefore, one needs to be careful when drawing conclusions on the fate of the $U(1)_A$ anomaly solely based on the observation of $\chi_{\rm top}$. 
 
There are several ways to improve our approach that may help overcome some of the issues, which were not treated in the present work. First, we had to introduce an IR cutoff to avoid singular behaviors of the flow equations stemming from the singularity of the bosonic occupation probability at low energy. This shortcoming is partly due to our approximation, where the effective masses of the fluctuations appearing in the flow equations are estimated solely by the coefficients of the quadratic terms in the fields. Therefore, the aforementioned singularities can be avoided, if one does not rely on a Taylor expansion of the effective potential ($V$), but treat it non-perturbatively, as done in Ref.~\cite{Fejos:2016hbp}, where no ansatz is assumed for the shape of $V$. Second, the anomaly enhancement at finite $T$ and $\mu_q$, which is one of our main results, is reliable only at comparably small temperatures and densities, i.e., $T\lesssim T_{\rm pc}$ and $ \mu_q\lesssim \mathring{M}_\pi$, since our analysis takes into account only hadronic fluctuations. At higher $T$ and $\mu_q$, the inclusion of quark excitations is necessary. 
Such an analysis can be done by adopting, e.g., the (Polyakov-)quark-meson-diquark model to QC$_2$D~\cite{Strodthoff:2011tz,Strodthoff:2013cua}, but this is beyond the scope of the present work and is left for future studies. Third, in the present analysis we have adopted the LPA with an expectation that the wave function renormalizations do not play a significant role in the medium. However, (momentum-dependent) wave function renormalization constants can play a central role in discussing peculiar energy dispersion relations, which may lead to, e.g., moat regimes~\cite{Rennecke:2023xhc,Fu:2024rto,Rennecke:2025kub}. We leave the examination of the renormalization constants and exploration of the moat regime for future work.
 
 
\section*{Acknowledgments}

G.F. was supported by the Hungarian National Research, Development, and Innovation Fund under Project No. FK142594. D.S. was supported by Grants-in-Aid for Scientific Research No. 23K03377, No. 23H05439 and No. 25K17386, from Japan Society for the Promotion of Science. D.S. thanks the KMI/FlaP Overseas Dispatch Program for supporting his stay at E\"{o}tv\"{o}s Lor\'{a}nd University. D.S. also thanks E\"{o}tv\"{o}s Lor\'{a}nd University for providing a comfortable research environment.

\appendix

\section{Derivation of flow equations}
\label{sec:FlowEquation}

Here, we illustrate our strategy to derive flow equations for the couplings, $m_{M}^2$, $m_B^2$, $\cdots$, from Eq.~(\ref{RHSExpand}), based on the LPA.

Let us concentrate on the ${\cal O}(\phi^2)$ piece in Eq.~(\ref{RHSExpand}), as a demonstration:
\begin{eqnarray}
\tilde{\partial}_k\Gamma_{k(2)} \equiv-  \frac{\tilde{\partial}_k}{2}T\sum_n\int_{\bm p}{\rm tr}\Big[D_{0,k}V_{(2)}''\Big] \ .
\end{eqnarray}
First, taking the second derivative of $V$ with respect to all fields $\phi \ni\{{\cal S}^a,{\cal P}^a\}$, one obtains the $12\times 12$ matrix $V''$. Then separating the ${\cal O}(\phi^0)$ piece, $V_{(0)}''$, from the ${\cal O}(\phi^2)$ one, $V_{(2)}''$, one gets the leading order, field independent propagator, $D_{0,k}$: see Eqs.~(\ref{DMeson1}),~(\ref{DMeson2}),~(\ref{DBaryon1}) and~(\ref{DBaryon2}). More explicitly, $\tilde{\partial}_k\Gamma_{k(2)}$ takes the form of
\begin{eqnarray}
&& \tilde{\partial}_k\Gamma_{k(2)} \nonumber\\
&=& -\frac{\tilde{\partial}_k}{2}\int_{\bm p}\Bigg[\frac{\sum_{a=1}^3V_{{\cal P}^a{\cal P}^a}''^{(2)}+V_{{\cal S}^0{\cal S}^0}''^{(2)}}{2E_{k,M+}}G(E_{k,M+},E_{k,M+}) \nonumber\\
&& + \frac{\sum_{a=1}^3V_{{\cal S}^a{\cal S}^a}''^{(2)}+V_{{\cal P}^0{\cal P}^0}''^{(2)}}{2E_{k,M-}}G(E_{k,M-},E_{k,M-}) \nonumber\\
&& + \frac{V_{{\cal P}^4{\cal P}^4}''^{(2)}+V_{{\cal P}^5{\cal P}^5}''^{(2)}}{2E_{k,B+}} G(E_{k,B+}-2\mu_q,E_{k,B+}+2\mu_q)\nonumber\\
&& + \frac{V_{{\cal S}^4{\cal S}^4}''^{(2)}+V_{{\cal S}^5{\cal S}^5}''^{(2)}}{2E_{k,B-}} G(E_{k,B-}-2\mu_q,E_{k,B-}+2\mu_q) \Bigg]  \nonumber\\
\label{Omega2}
\end{eqnarray}
after the Matsubara summation, where $V_{\phi\phi}''^{(2)} = \partial^2 V/\partial \phi\partial\phi$, and 
\begin{eqnarray}
G(\epsilon_1,\epsilon_2) \equiv  1 +f_B(\epsilon_1) + f_B(\epsilon_2)\ ,
\end{eqnarray}
 with the help of Eq.~(\ref{MatsubaraF}) and the identity $f_B(-\epsilon) = -1-f_B(\epsilon)$. We note that the quadratic terms of the $12$ fields are contained in $V_{\phi\phi}''^{(2)}$.

Based on all symmetries of QC$_2$D at finite $\mu_q$, in homogeneous backgrounds the integrand in Eq.~(\ref{Omega2}) can be collected into a form of 
\begin{eqnarray}
\tilde{\partial}_k\Gamma_{k(2)} &=& {\cal F}_{m_M^2,k}I_{\rm AF1}^M  + {\cal F}_{m_B^2,k} I_{\rm AF1}^B  \nonumber\\
&&+  {\cal F}_{a_M,k} I_{\rm A1}^M +  {\cal F}_{a_B,k} I_{\rm A1}^B\ ,\label{Omega2Form}
\end{eqnarray}
with the invariants being
\begin{eqnarray}
I_{\rm AF1}^M &\equiv& {\rm tr}\left[\Sigma_M^\dagger\Sigma_M\right] \ ,  \nonumber\\
I_{\rm AF1}^B &\equiv& {\rm tr}\left[\Sigma_{B_R}^\dagger\Sigma_{B_R} + \Sigma_{B_L}^\dagger\Sigma_{B_L}\right] \ , \nonumber\\
I_{\rm A1}^M &\equiv& {\rm det}\Sigma_M+{\rm det}\Sigma_M^\dagger\ , \nonumber\\
I_{\rm A1}^B &\equiv&  {\rm tr}\left[\Sigma_{B_R}^\dagger\Sigma_{B_L} + \Sigma_{B_L}^\dagger\Sigma_{B_R}\right]\ ,
\end{eqnarray}
as we are working in the LPA. However, Eq.~(\ref{Omega2}) looks rather complicated and it is hard to see how the invariants actually emerge. Since in the end the terms in the r.h.s. of (\ref{Omega2}) must be combined into an expression compatible with Eq.~(\ref{Omega2Form}), it is sufficient to work with four independent fields to find $ {\cal F}_{m_M^2,k}$, $ {\cal F}_{m_B^2,k}$, $ {\cal F}_{a_M,k}$ and $ {\cal F}_{a_B,k}$. For example, when leaving ${\cal S}^0$, ${\cal P}^0$, ${\cal S}^4$, and ${\cal P}^4$, one can formally get
\begin{eqnarray}
&& {\cal P}_0 = \frac{1}{2}\sqrt{I_{\rm AF1}^M- I_{\rm A1}^M}\ , \ \ {\cal S}_0 = \frac{1}{2}\sqrt{I_{\rm AF1}^M+ I_{\rm A1}^M}\ , \nonumber\\
&& {\cal P}_4 = \frac{1}{2\sqrt{2}}\sqrt{I_{\rm AF1}^B + I_{\rm A1}^B}\ , \ \ {\cal S}_4 = \frac{1}{2\sqrt{2}}\sqrt{I_{\rm AF1}^B - I_{\rm A1}^B}\ . \nonumber\\
\end{eqnarray}
Inserting these expressions into $V_{\phi\phi}''^{(2)}$ in Eq.~(\ref{Omega2}) immediately provides the coefficients $ {\cal F}_{m_M^2,k}$, $ {\cal F}_{m_B^2,k}$, $ {\cal F}_{a_M,k}$ and $ {\cal F}_{a_B,k}$, which yields the flow equations for $m_M^2$, $m_B^2$, $a_M$ and $a_B$:
\begin{eqnarray}
&& \partial_k m_M^2 = {\cal F}_{m_M^2,k}\ , \ \ \partial_k m_B^2 = {\cal F}_{m_B^2,k} \ ,\nonumber\\
&&   \partial_k a_M = {\cal F}_{a_M,k}\ , \ \ \partial_k a_B = {\cal F}_{a_B,k}\ .
\end{eqnarray}

It should be noted that the ${\bm p}$ integrations can be easily carried out using ($\alpha=M\pm, B\pm$)
 \begin{eqnarray}
\tilde{\partial}_k\int_{\bm p} {\mathscr F}(E_{k,\alpha})&=& \int_{\bm p} 2k\, \theta(k^2-{\bm p}^2)\frac{\partial}{\partial R_k}{\mathscr F}(E_{k,\alpha}) \nonumber\\
&=& \frac{k^4}{3\pi^2}\frac{k}{E_{k,\alpha}}\frac{\partial{\mathscr F}(E_{k,\alpha})}{\partial E_{k,\alpha}}\Bigg|_{E_{k,\alpha} = \sqrt{k^2+(m_\alpha^{\rm eff})^2}}\, . \nonumber\\
\end{eqnarray}

In a similar manner, flow equations for the remaining four-point couplings: $\lambda_{M1}$, $\lambda_{M2}$, $\lambda_{B1}$, $\lambda_{B2}$, $\gamma_1$, $\gamma_2$, $c_{M1}$, $c_{M2}$, $c_{B1}$, $c_{B2}$, $d_1$, $d_2$ and $d_3$ can be derived.

\section{Potential with $\mu_q=0$}
\label{sec:PotentialSU4}

Here, we derive relationships between the couplings, $m_M^2, m_B^2,\cdots$, which hold for vanishing chemical potential, $\mu_q=0$.

The potential in Eq.~(\ref{VSum}) is constructed such that it preserves all symmetries of QC$_2$D at finite $\mu_q$, $SU(2)_L\times SU(2)_R$ chiral, $U(1)_B$, parity and time-reversal symmetries. However, for vanishing $\mu_q$, QC$_2$D possesses an extended set of symmetries, i.e., the Pauli-G\"{u}rsey $SU(4)$, parity, time-reversal and charge-conjugation symmetries. In this limit, a general LSM potential $\tilde{V}$ up to ${\cal O}(\Sigma^4)$ was derived to be \cite{Fejos:2025oxi}
\begin{eqnarray}
\tilde{V} \equiv \tilde{m}^2\tilde{I}_1+ \tilde{\lambda}_1\tilde{I}_1^2+\tilde{\lambda}_2\tilde{I}_2 + \tilde{a}\tilde{I}_A + \tilde{c}_1\tilde{I}_A' + \tilde{c}_2\tilde{I}_1\tilde{I}_A\ , \label{SU4Potential}
\end{eqnarray}
where we have defined the following invariants:
\begin{eqnarray}
&& \tilde{I}_1 \equiv {\rm tr}[\Sigma^\dagger\Sigma]\ , \ \ \tilde{I}_A \equiv \frac{1}{2}{\rm tr}\left[\tilde{\Sigma}\Sigma+\tilde{\Sigma}^\dagger\Sigma^\dagger\right] \ , \nonumber\\
&& \tilde{I}'_A \equiv \tilde{I}_A^2-\frac{1}{2}{\rm tr}\left[\tilde{\Sigma}\Sigma\right]{\rm tr}\left[\tilde{\Sigma}^\dagger\Sigma^\dagger\right]  = 4\left({\rm det}\Sigma + {\rm det}\Sigma^\dagger\right)\ , \nonumber\\
&& \tilde{I}_2 \equiv  {\rm tr}\left[\left(\Sigma^\dagger\Sigma-\frac{{1}}{4}{\rm tr}[\Sigma^\dagger\Sigma]\right)^2\right]\ ,
\end{eqnarray}
see the definition of $\Sigma$ in Eq.~(\ref{SigmaDef}). Note that we have introduced the $\tilde{I}_A'$ invariant instead of simply using $\tilde{I}^2_A$, as done in Ref.~\cite{Fejos:2025oxi}, the former being more transparent as the anomalous and non-anomalous contributions are completely separated. That is, $\tilde{m}^2$, $\tilde{\lambda}_1$ and $\tilde{\lambda}_2$ terms are non-anomalous, while $\tilde{a}$, $\tilde{c}_1$ and $\tilde{c}_2$ are irreducibly anomalous. Here, one can easily confirm the following identities between invariants:
\begin{eqnarray}
\tilde{I}_1 = \frac{1}{8}{\rm tr}\left[\Sigma_{B_R}^\dagger\Sigma_{B_R} + \Sigma_{B_L}^\dagger\Sigma_{B_L}\right] + \frac{1}{4}{\rm tr}\left[\Sigma_M^\dagger\Sigma_M\right] \, ,
\end{eqnarray}
\begin{eqnarray}
\tilde{I}_A =  \frac{1}{4}({\rm det}\Sigma_M + {\rm det}\Sigma_M^\dagger) + \frac{1}{8}{\rm tr}\Big[\Sigma_{B_R}^\dagger\Sigma_{B_L}+\Sigma_{B_L}^\dagger\Sigma_{B_R}\Big] \, , \nonumber\\
\end{eqnarray}
\begin{eqnarray}
\tilde{I}_2 &=& \frac{1}{32}{\rm tr}\left[(\Sigma_M^\dagger\Sigma_M)^2\right] - \frac{1}{64}\left({\rm tr}[\Sigma_M^\dagger\Sigma_M]\right)^2 \nonumber\\
&& +  \frac{1}{256}\left({\rm tr}\left[\Sigma_{B_R}^\dagger\Sigma_{B_R} - \Sigma_{B_L}^\dagger\Sigma_{B_L}\right]\right)^2  \nonumber\\
&& + \frac{1}{64}{\rm tr}\Big[\Sigma_{B_R}^\dagger\Sigma_{B_R}+\Sigma_{B_L}^\dagger\Sigma_{B_L}\Big] {\rm tr}\left[\Sigma_{M}^\dagger\Sigma_{M} \right] \nonumber\\
&&- \frac{1}{32} \left( {\rm tr}\Big[\Sigma_{B_R}^\dagger\Sigma_{B_L}\Big] {\rm det}\Sigma_M^\dagger + {\rm tr}\Big[\Sigma_{B_L}^\dagger\Sigma_{B_R}\Big] {\rm det}\Sigma_M   \right) \, , 
\nonumber\\
\end{eqnarray}
\begin{eqnarray}
\tilde{I}_1^2 &=& \frac{1}{64}\left({\rm tr}\left[\Sigma_{B_R}^\dagger\Sigma_{B_R}  + \Sigma_{B_L}^\dagger\Sigma_{B_L}\right]\right)^2 \nonumber\\
&& + \frac{1}{16}{\rm tr}\left[\Sigma_M^\dagger\Sigma_M\right] {\rm tr}\left[\Sigma_{B_R}^\dagger\Sigma_{B_R} + \Sigma_{B_L}^\dagger\Sigma_{B_L}\right] \nonumber\\
&& + \frac{1}{16}\left({\rm tr}\left[\Sigma_M^\dagger\Sigma_M\right]\right)^2 \ ,
\end{eqnarray}
\begin{eqnarray}
&& \tilde{I}_1\tilde{I}_A =  \frac{1}{16}{\rm tr}\left[\Sigma_M^\dagger\Sigma_M\right] \left({\rm det}\Sigma_M + {\rm det}\Sigma_M^\dagger\right) \nonumber\\
&& + \frac{1}{32}{\rm tr}\left[\Sigma_M^\dagger\Sigma_M\right] {\rm tr}\Big[\Sigma_{B_R}^\dagger\Sigma_{B_L}+\Sigma_{B_L}^\dagger\Sigma_{B_R}\Big] \nonumber\\
&& + \frac{1}{32} {\rm tr}\Big[\Sigma_{B_R}^\dagger\Sigma_{B_R}+\Sigma_{B_L}^\dagger\Sigma_{B_L}\Big] \left({\rm det}\Sigma_M + {\rm det}\Sigma_M^\dagger\right) \nonumber\\
&& + \frac{1}{64} {\rm tr}\Big[\Sigma_{B_R}^\dagger\Sigma_{B_R}+\Sigma_{B_L}^\dagger\Sigma_{B_L}\Big] {\rm tr}\Big[\Sigma_{B_R}^\dagger\Sigma_{B_L}+\Sigma_{B_L}^\dagger\Sigma_{B_R}\Big]\ , \nonumber\\
\end{eqnarray}
and
\begin{eqnarray}
\tilde{I}_A' &=& \frac{1}{16} \left(({\rm det}\Sigma_M)^2 + ({\rm det}\Sigma_M^\dagger)^2\right) \nonumber\\
&& + \frac{1}{16}\left({\rm tr}[\Sigma_{B_R}^\dagger\Sigma_{B_L}]{\rm det}\Sigma_M +{\rm tr}[\Sigma_{B_L}^\dagger\Sigma_{B_R}] {\rm det}\Sigma_M^\dagger\right) \nonumber\\
&& + \frac{1}{64}\left(({\rm tr}[\Sigma_{B_R}^\dagger\Sigma_{B_L}])^2+({\rm tr}[\Sigma_{B_L}^\dagger\Sigma_{B_R}])^2 \right) \ .
\end{eqnarray}
Thus, the $SU(4)$ symmetric potential $\tilde{V}$ can be rewritten into
\begin{widetext}
\begin{eqnarray}
\tilde{V} &=& \frac{\tilde{m}^2}{4}{\rm tr}\left[\Sigma_M^\dagger\Sigma_M\right] +\frac{\tilde{m}^2}{8}{\rm tr}\left[\Sigma_{B_R}^\dagger\Sigma_{B_R} + \Sigma_{B_L}^\dagger\Sigma_{B_L}\right]  + \frac{\tilde{\lambda}_1}{16}\left({\rm tr}\left[\Sigma_M^\dagger\Sigma_M\right]\right)^2+ \frac{\tilde{\lambda}_2}{32} {\rm tr}\left[\left(\Sigma_M^\dagger\Sigma_M - \frac{1}{2}{\rm tr}[\Sigma_M^\dagger\Sigma_M]\right)^2\right] \nonumber\\
&+& \frac{\tilde{\lambda}_1}{64}\left({\rm tr}\left[\Sigma_{B_R}^\dagger\Sigma_{B_R} + \Sigma_{B_L}^\dagger\Sigma_{B_L}\right]\right)^2 +  \frac{\tilde{\lambda}_2}{256}\left({\rm tr}\left[\Sigma_{B_R}^\dagger\Sigma_{B_R} - \Sigma_{B_L}^\dagger\Sigma_{B_L}\right]\right)^2  \nonumber\\
&+&  \frac{4\tilde{\lambda}_1+\tilde{\lambda}_2}{64}{\rm tr}\left[\Sigma_M^\dagger\Sigma_M\right] {\rm tr}\left[\Sigma_{B_R}^\dagger\Sigma_{B_R} + \Sigma_{B_L}^\dagger\Sigma_{B_L}\right]  -\frac{\tilde{\lambda}_2}{32} \left({\rm tr}\left[\Sigma_{B_R}^\dagger\Sigma_{B_L} \right]{\rm det}\Sigma^\dagger_M + {\rm tr}\left[ \Sigma_{B_L}^\dagger\Sigma_{B_R}\right]{\rm det}\Sigma_M \right) \nonumber\\
&+& \frac{\tilde{a}}{4}({\rm det}\Sigma_M + {\rm det}\Sigma_M^\dagger) + \frac{\tilde{a}}{8}{\rm tr}\Big[\Sigma_{B_R}^\dagger\Sigma_{B_L}+\Sigma_{B_L}^\dagger\Sigma_{B_R}\Big]  \nonumber\\
&+&  \frac{\tilde{c}_1}{16} \left(({\rm det}\Sigma_M)^2 + ({\rm det}\Sigma_M^\dagger)^2\right) + \frac{\tilde{c_2}}{16}{\rm tr}\left[\Sigma_M^\dagger\Sigma_M\right] \left({\rm det}\Sigma_M  + {\rm det}\Sigma_M^\dagger\right) \nonumber\\
&+& \frac{\tilde{c}_1}{64}\left(({\rm tr}[\Sigma_{B_R}^\dagger\Sigma_{B_L}])^2+({\rm tr}[\Sigma_{B_L}^\dagger\Sigma_{B_R}])^2 \right) +  \frac{\tilde{c_2}}{64} {\rm tr}\Big[\Sigma_{B_R}^\dagger\Sigma_{B_R}+\Sigma_{B_L}^\dagger\Sigma_{B_L}\Big] {\rm tr}\Big[\Sigma_{B_R}^\dagger\Sigma_{B_L}+\Sigma_{B_L}^\dagger\Sigma_{B_R}\Big] \nonumber\\ 
&+& \frac{\tilde{c_2}}{32}{\rm tr}\left[\Sigma_M^\dagger\Sigma_M\right] {\rm tr}\Big[\Sigma_{B_R}^\dagger\Sigma_{B_L}+\Sigma_{B_L}^\dagger\Sigma_{B_R}\Big] + \frac{\tilde{c_2}}{32} {\rm tr}\Big[\Sigma_{B_R}^\dagger\Sigma_{B_R}+\Sigma_{B_L}^\dagger\Sigma_{B_L}\Big] \left({\rm det}\Sigma_M + {\rm det}\Sigma_M^\dagger\right)\nonumber\\
&+&  \frac{\tilde{c}_1}{16}\left({\rm tr}\left[\Sigma_{B_R}^\dagger\Sigma_{B_L} \right]{\rm det}\Sigma_M + {\rm tr}\left[ \Sigma_{B_L}^\dagger\Sigma_{B_R}\right]{\rm det}\Sigma^\dagger_M \right)  \ ,\label{FullPotential}
\end{eqnarray}
\end{widetext}
in terms of the mesonic and baryonic $2\times2$ matrices: $\Sigma_M$, $\Sigma_{B_R}$ and $\Sigma_{B_L}$. By comparing (\ref{FullPotential}) with the potentials~(\ref{VAFGeneral1}) and~(\ref{VAGeneral1}), one is immediately led to the following $SU(4)$ symmetric relationships between the couplings:
\begin{eqnarray}
&& m_M^2=2m_B^2= \frac{\tilde{m}^2}{4}\ , \ \ \lambda_{M1} = 4 \lambda_{B1} = \gamma_1+\frac{\gamma_2}{2}= \frac{\tilde{\lambda}_1}{16}\ ,  \nonumber\\
&& \lambda_{M2} = 8\lambda_{B2} =-\gamma_2= \frac{\tilde{\lambda}_2}{32}\ , \ \ a_M=2a_B = \frac{\tilde{a}}{4}\ , \nonumber\\
&& c_{M1} = 4c_{B1} = d_3 = \frac{\tilde{c}_1}{16}\ , \nonumber\\
&&  c_{M2} = 4c_{B2} = 2d_1=2d_2=\frac{\tilde{c}_2}{16}\ . \label{SU4LimitParameter}
\end{eqnarray}

\section{Determination of $k_{\rm IR}$}
\label{sec:KIRDetermination}

The r.h.s. of flow equations contains the Bose-Einstein distribution function of the form $f_B(\sqrt{k^2+(m_{B\pm}^{\rm eff})^2}-2\mu_q)$, which can lead to a singular behavior when $k \sim \sqrt{4\mu_q^2-(m_{B\pm}^{\rm eff})^2}$, which leads to the breakdown of our approximation. In order to avoid this singularity, we employ an IR cutoff, $k_{\rm IR}$. Here, we discuss how to determine this IR cutoff.

\begin{figure}[t]
\centering
\hspace*{-0.2cm} 
\includegraphics*[scale=0.6]{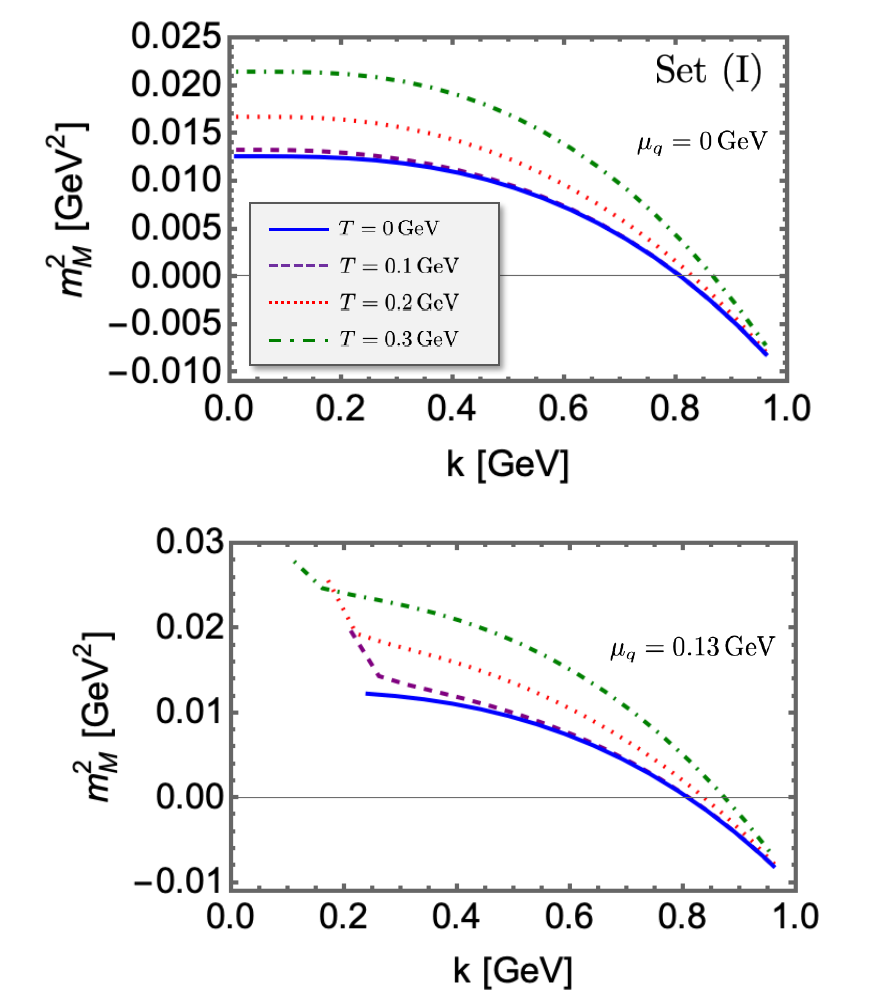}
\caption{Scale ($k$) dependence of the parameter $m_M^2$ at several temperatures with $\mu_q=0$ GeV (top) and $\mu_q=0.13$ GeV (bottom), for the parameter set (I).}
\label{fig:KDepSet1}
\end{figure}

\begin{figure}[t]
\centering
\hspace*{-0.2cm} 
\includegraphics*[scale=0.6]{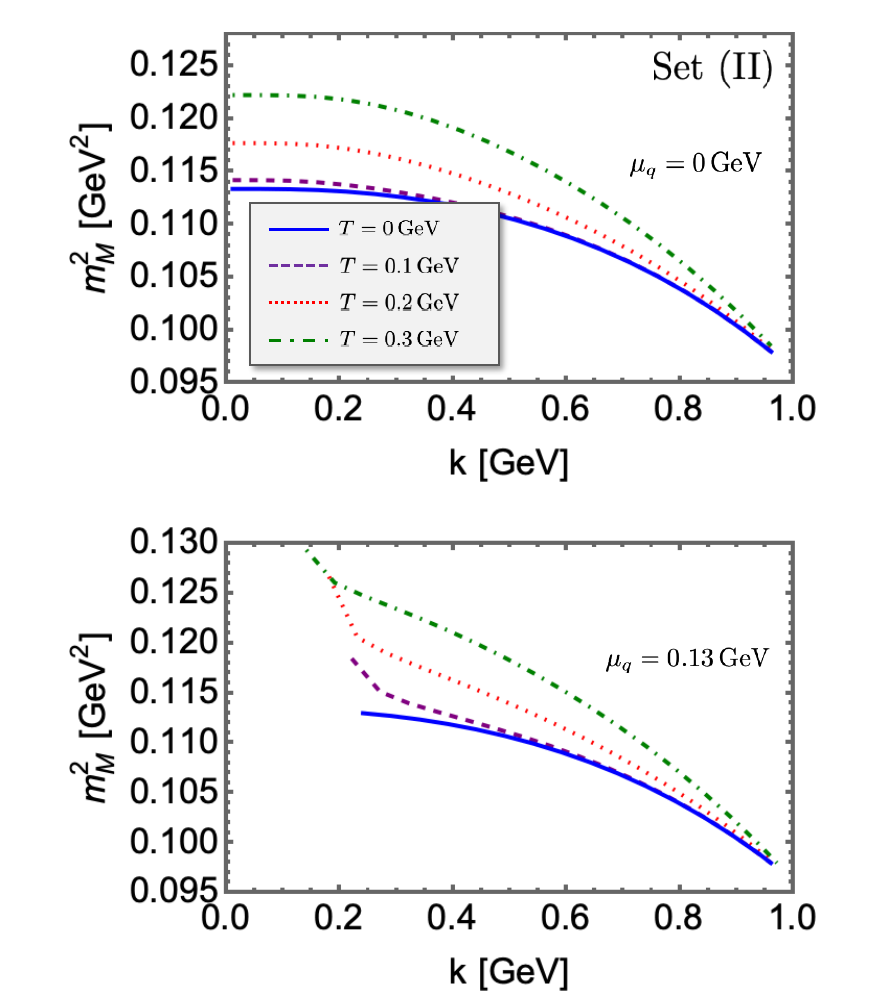}
\caption{Scale ($k$) dependence of the parameter $m_M^2$ at several temperatures with $\mu_q=0$ GeV (top) and $\mu_q=0.13$ GeV (bottom), for the parameter set (II).}
\label{fig:KDepSet2}
\end{figure}

The breakdown of the present approximation, i.e., divergence of the r.h.s. of the flow equations, are reflected by the convergence of the couplings: $m_M^2$, $m_B^2$, $\cdots$. Since there are $17$ couplings, we only focus on $m_M^2$ as a representative. Depicted in Figs.~\ref{fig:KDepSet1} and~\ref{fig:KDepSet2} are $k$ dependencies of $m_M^2$ at $\mu_q=0,0.13$ GeV for several temperatures, with the parameter set (I) and (II), respectively. The top panels indicate that at $\mu_q=0$ GeV, the value for $m_M^2$ converges well for $k\lesssim 0.3$ GeV. At $\mu_q=0.13$ GeV, however, abrupt changes of $m_M^2$ for $k\lesssim 0.3$ GeV are derived, as shown in the bottom panels, meaning the breakdown of our approximation. The curves vanish at which the momentum hits $k = \sqrt{4\mu_q^2-(m_{B\pm}^{\rm eff})^2}$, leading to divergences of the bosonic occupation probability. We note that the abrupt change is absent when $T=0$ GeV, since in this case the distribution function $f_B(\sqrt{k^2+(m_{B\pm}^{\rm eff})^2}-2\mu_q)$ is always zero except for $k=\sqrt{4\mu_q^2-(m_{B\pm}^{\rm eff})^2}$.

Adopting a common $k_{\rm IR}$ value at any $T$ and $\mu_q$ is reasonable to study modifications in medium in a self-consistent way. Therefore, we employ $k_{\rm IR} = 0.3$ GeV as a suitable choice for $T\lesssim 0.3$ GeV and $\mu_q\lesssim \mathring{M}_\pi$, as seen from Figs.~\ref{fig:KDepSet1} and~\ref{fig:KDepSet2}. It should be noted that at higher temperatures, the convergence gets worse, thus, we have to keep in mind that all the estimates with $k_{\rm IR}=0.3$ GeV at higher temperatures underestimate the ``exact" results. These properties are similar for the other coefficients as well.

\section{Hadron mass formulas}
\label{sec:HadronMassApp}

Here we show hadron mass formulas based on our present LSM. We note that we do not attach the $k$ subscript anywhere, for simplicity. 

In the presence of chiral and diquark condensates at finite $\mu_q$, the scalar fields ${\cal S}^0$ and ${\cal P}^5$ may acquire vacuum expectation values: $\sigma_0 \equiv \langle {\cal S}^0\rangle$ and $\Delta \equiv \langle{\cal P}^5\rangle$. Upon these mean fields, pion and $a_0$ masses are straightforwardly read off as the coefficients of the respective quadratic terms:
\begin{eqnarray}
M_\pi^2 &=& 4(m_M^2+a_M) + 8(c_{M1}+2c_{M2}+2\lambda_{M1})\sigma_0^2 \nonumber\\
&& + 8(2d_1+2d_2+d_3+2\gamma_1+\gamma_2)\Delta^2 \ , \nonumber\\
M_{a_0}^2 &=& 4(m_M^2-a_M) + 8(-c_{M1}+2\lambda_{M1}+2\lambda_{M2})\sigma_0^2 \nonumber\\
&& + 8(2d_1-2d_2-d_3+2\gamma_1-\gamma_2)\Delta^2 \ . \label{PiA0Mass}
\end{eqnarray}

Meanwhile, $\sigma$, $B$ and $\bar{B}$ or $\eta$, $B'$ and $\bar{B}'$ mix in the superfluid phase due to the violation of $U(1)_B$ symmetry. Their masses are evaluated by the poles of the corresponding propagator matrices at the rest frame. The propagator inverses read
\begin{eqnarray}
&& D_{\sigma B\bar{B}}^{-1} \nonumber\\
&=& \left(
\begin{array}{ccc}
p^2-V''_{{\cal S}^0{\cal S}^0} & 0 & -V''_{{\cal S}^0{\cal P}^5} \\
0 & p^2+4\mu_q^2-V''_{{\cal P}^4{\cal P}^4} &-4i\mu_qp_0\\
-V''_{{\cal S}^0{\cal P}^5} & 4i\mu_q p_0 & p^2+4\mu_q^2-V''_{{\cal P}^5{\cal P}^5} \\
\end{array}
\right)\, , \nonumber\\
\end{eqnarray}
and
\begin{eqnarray}
&& D_{\eta B'\bar{B}'}^{-1} \nonumber\\
&=& \left(
\begin{array}{ccc}
p^2-V''_{{\cal P}^0{\cal P}^0} & 0 & -V''_{{\cal P}^0{\cal S}^5} \\
0 & p^2+4\mu_q^2-V''_{{\cal S}^4{\cal S}^4} &-4i\mu_qp_0\\
-V''_{{\cal P}^0{\cal S}^5} & 4i\mu_q p_0 & p^2+4\mu_q^2-V''_{{\cal S}^5{\cal S}^5} \\
\end{array}
\right)\, , \nonumber\\ \label{DEtaBBp}
\end{eqnarray}
with
\begin{eqnarray}
V''_{{\cal S}_0{\cal S}_0} &=& 4(m_M^2+a_M) + 24(c_{M1}+2c_{M2}+2\lambda_{M1})\sigma_0^2 \nonumber\\
&& + 8(2d_1+2d_2+d_3+2\gamma_1+\gamma_2)\Delta^2\ , \nonumber\\
V''_{{\cal P}_4{\cal P}_4} &=& 8(m_B^2+a_B) + 8(2d_1+2d_2+d_3+2\gamma_1+\gamma_2)\sigma_0^2 \nonumber\\
&&  + 32(c_{B1}+2c_{B2}+2\lambda_{B1})\Delta^2\ , \nonumber\\
V''_{{\cal P}_5{\cal P}_5} &=& 8(m_B^2+a_B) + 8(2d_1+2d_2+d_3+2\gamma_1+\gamma_2)\sigma_0^2 \nonumber\\
&& + 96(c_{B1}+2c_{B2}+2\lambda_{B1})\Delta^2\ , \nonumber\\
V''_{{\cal S}_0{\cal P}_5} &=& 16(2d_1+2d_2+d_3+2\gamma_1+\gamma_2)\Delta\sigma_0 \ , 
\end{eqnarray}
and
\begin{eqnarray}
V''_{{\cal P}_0{\cal P}_0} &=& 4(m_M^2-a_M) + 8(-3c_{M1}+2\lambda_{M1})\sigma_0^2 \nonumber\\
&& + 8(2d_1-2d_2-d_3+2\gamma_1-\gamma_2)\Delta^2\ , \nonumber\\
V''_{{\cal S}_4{\cal S}_4} &=& 8(m_B^2-a_B) - 8(2d_1-2d_2 + d_3-2\gamma_1+\gamma_2)\sigma_0^2 \nonumber\\
&& + 32(-c_{B1}+2\lambda_{B1} +4 \lambda_{B2})\Delta^2\ , \nonumber\\
V''_{{\cal S}_5{\cal S}_5} &=& 8(m_B^2-a_B) - 8(2d_1- 2d_2 + d_3 - 2\gamma_1+\gamma_2)\sigma_0^2 \nonumber\\
&& + 32(-3c_{B1}+2\lambda_{B1})\Delta^2\ , \nonumber\\
V''_{{\cal P}_0{\cal S}_5} &=& 16(d_3-\gamma_2)\Delta\sigma_0 .
\end{eqnarray}

In the hadronic phase, where $\Delta=0$, the hadron masses are simply given by
\begin{eqnarray}
M_\pi^{\rm (H)} &=& \sqrt{4(m_M^2+a_M) + 8(c_{M1}+2c_{M2}+2\lambda_{M1})\sigma_0^2 } \ , \nonumber\\
M_{a_0}^{\rm (H)} &=& \sqrt{4(m_M^2-a_M) + 8(-c_{M1}+2\lambda_{M1}+2\lambda_{M2})\sigma_0^2}  \ , \nonumber\\
\end{eqnarray}
\begin{eqnarray}
M_\sigma^{\rm (H)} &=&\sqrt{ 4(m_M^2+a_M) + 24(c_{M1}+2c_{M2}+2\lambda_{M1})\sigma_0^2} \ ,\nonumber\\
M_B^{\rm (H)} &=& \sqrt{ 8(m_B^2+a_B) + 8(2d_1+2d_2+d_3+2\gamma_1+\gamma_2)\sigma_0^2} \nonumber\\
&& -2\mu_q\ , \nonumber\\
M_{\bar{B}}^{\rm (H)} &=& \sqrt{ 8(m_B^2+a_B) + 8(2d_1+2d_2+d_3+2\gamma_1+\gamma_2)\sigma_0^2} \nonumber\\
&& + 2\mu_q\ ,  \label{MassHadron1}
\end{eqnarray}
and
\begin{eqnarray}
M_\eta^{\rm (H)}&=&  \sqrt{4(m_M^2-a_M) + 8(-3c_{M1}+2\lambda_{M1})\sigma_0^2 } \ ,\nonumber\\
M_{B'}^{\rm (H)} &=& \sqrt{ 8(m_B^2-a_B) - 8(2d_1-2d_2 + d_3-2\gamma_1+\gamma_2)\sigma_0^2} \nonumber\\
&& - 2\mu_q\ , \nonumber\\
M_{\bar{B}'}^{\rm (H)} &=& \sqrt{ 8(m_B^2-a_B) - 8(2d_1-2d_2 + d_3-2\gamma_1+\gamma_2)\sigma_0^2} \nonumber\\
&& + 2\mu_q \ .  \nonumber\\
\end{eqnarray}

\bibliography{reference}

\end{document}